\documentclass[12pt]{article}
\usepackage[authoryear,round,semicolon]{natbib}
\usepackage{authblk}
\usepackage{amsmath}
\usepackage{amstext}
\usepackage{floatrow}
\usepackage{dsfont}
\usepackage{changepage} 
\usepackage{setspace}
\usepackage[hang, flushmargin]{footmisc}
\usepackage[hidelinks]{hyperref}
\usepackage{footnotebackref}
\usepackage{adjustbox}  
\usepackage{rotating}
\usepackage{amssymb}
\usepackage{graphicx}
\usepackage[english]{babel}
\usepackage[autostyle]{csquotes}
\usepackage{geometry}
\usepackage{makecell}
\usepackage{array}
\usepackage{lscape}
\usepackage{tabularx}
\usepackage{bbm}
\usepackage{caption}
\usepackage{ragged2e}
\usepackage{tikz}
\usepackage{amsmath}
\usepackage{subcaption}
\usepackage{siunitx}
\usepackage{array}
\usepackage{threeparttable}
\newcolumntype{C}[1]{>{\centering\let\newline\\\arraybackslash\hspace{0pt}}m{#1}}
\DeclareGraphicsExtensions{.pdf,.png,.jpg, .eps}
\newgeometry{left=3.5cm,right=3.5cm,bottom=4cm}

\begin{document}

\title{When Networks Mislead: How Partisan Communication Undermines Democratic Decision-Making}

\author[1]{Hsuan-Wei Lee}
\author[2]{Po-Kang Hsiao}
\affil[1]{College of Health, Lehigh University, USA}
\affil[2]{School of Business, University of Wisconsin-Madison, USA}
\affil[*]{Corresponding author: hsl324@lehigh.edu}

\date{}

\maketitle

\begin{abstract}
Democratic societies increasingly rely on communication networks to aggregate citizen preferences and information, yet these same networks can systematically mislead voters under certain conditions. We introduce an agent-based model that captures two rival forces in partisan networks: honest noise filtering that lifts accuracy and strategic bluffing that embeds bias. Extensive simulations show that communication architecture shapes voting accuracy more than any individual-level trait. When candidate quality gaps are moderate, partisan bluffing overpowers honest signals and steers supporters of weaker contenders into collective error. However, positioning independents in central network roles serves as an epistemic circuit breaker, preventing echo chambers from spiraling toward systematic error. Counterintuitively, we discover that competitive elections with meaningful quality differences prove most vulnerable to collective delusion. Small numbers of extreme partisans can contaminate entire communities through cascading bias effects that persist across hundreds of communication rounds. Our findings challenge conventional wisdom about information aggregation in democracy and provide actionable insights for institutional design in an era of algorithmic filtering and social media polarization.

\end{abstract}


\newpage

\section{Introduction}

Democratic legitimacy rests on a fundamental assumption: that citizens possess the capacity to identify candidates who will best serve their interests and promote collective welfare through informed electoral choice. Yet recent evidence shows voters often back candidates who violate their own policy goals, reward poor performance, and form choices that diverge from objective measures of competence. This phenomenon has profound implications that extend far beyond individual voting decisions to threaten the epistemic foundations of democratic governance itself: if citizens cannot reliably distinguish between superior and inferior candidates, the theoretical justification for democratic rule—that collective choice aggregates dispersed information to produce superior outcomes—comes under fundamental strain. Digital media now amplify both collective learning and strategic manipulation, making the puzzle of voter error more urgent than ever \citep{biswas2014influence, ohme2018uncertain}.

Most scholarship treats voting accuracy as a function of personal knowledge, education, and cognitive bias. While extensive research has documented how political knowledge \citep{lau1997voting,lau2008exploration}, cognitive biases \citep{taber2006motivated}, and partisan media exposure \citep{levendusky2013partisan} affect individual voting decisions, and despite evidence that knowledgeable voters can sometimes make worse choices due to motivated reasoning \citep{hannon2022knowledgeable}, this approach overlooks how social networks fundamentally reshape the information environment through which citizens navigate electoral choices. This research tradition fundamentally mischaracterizes democratic choice by treating voters as isolated information processors rather than members of densely connected social networks that shape how evidence is gathered, interpreted, and transformed into political judgments. Because citizens now rely on online and offline networks for news, network structure increasingly drives electoral outcomes in ways that individual-level models miss. The rise of algorithmic curation on social media platforms has intensified this networked character of democratic communication, creating information environments where the structural properties of communication networks may matter more for electoral outcomes than individual voter characteristics.

This networked reality poses a fundamental theoretical puzzle that existing scholarship has yet to resolve: under what conditions do communication networks enhance versus undermine democratic decision-making? Social networks can play two opposite roles in political communication. They may facilitate beneficial information aggregation by enabling citizens to pool diverse perspectives, filter out idiosyncratic noise through repeated social interaction, and harness the statistical properties of large numbers to approach truth through collective deliberation—the optimistic vision underlying the Condorcet Jury Theorem \citep{austen1996information} and ``wisdom of crowds" theories \citep{surowiecki2005wisdom}. Alternatively, networks may systematically amplify bias and misinformation when strategic actors manipulate information flows for partisan advantage, when homophily creates isolated echo chambers that resist corrective information, or when social influence mechanisms produce cascades that carry entire communities away from accurate beliefs toward collectively harmful delusions. The same structural features that enable beneficial information aggregation can simultaneously serve as conduits for bias amplification that undermines democratic institutions' epistemic functions. Digital platforms have intensified these dynamics by creating environments where false information spreads faster and more broadly than truth \citep{vosoughi2018spread}, algorithmic filtering reduces exposure to ideologically diverse content \citep{bakshy2015exposure}, and social influence effects can mobilize political behavior across massive networks \citep{bond201261}. Recent computational research has demonstrated how network structure and agent behavior coevolve to shape collective outcomes in political contexts \citep{lee2018evolutionary, momennejad2022collective}, while digital environments introduce new sources of systematic bias through algorithmic actors that can artificially polarize discourse \citep{lu2024agents}.

Existing approaches to understanding networked political communication provide important insights but leave critical gaps that prevent comprehensive analysis of when networks help versus harm democratic performance. Formal models of information aggregation, building on the Condorcet tradition, typically assume unbiased communication between rational actors who share common interests in truth-seeking, thereby abstracting away from the strategic considerations and partisan identities that fundamentally shape real political discourse. Empirical studies reveal that political discussion networks significantly influence voting behavior \citep{huckfeldt1987networks,mcclurg2006electoral,ryan2010effects} and can either promote correct voting when networks provide clear signals \citep{richey2008social, sokhey2012social,watts2014influence} or undermine democratic decision-making when disagreement levels are high \citep{mutz2002cross,pattie2009conversation} or when networks become polarized \citep{butters2022polarized} and increasingly homophilic in their partisan composition. Yet we still lack clear evidence on how specific network features raise or lower collective accuracy. Our study speaks to this gap by linking measurable network metrics, such as degree centrality and clustering, to predicted shifts in correct-vote rates. Recent work has shown how social media activists can significantly influence voting patterns \citep{kofi2022impact} and how machine learning approaches can predict political preferences from network connections \citep{idan2019show}, yet the fundamental question of when networks help versus harm collective decision-making remains unresolved. Laboratory experiments on social learning, while methodologically rigorous, often employ artificial decision tasks that lack the partisan stakes and strategic complexity characterizing actual electoral contexts. \cite{ryan2011social} represents an important exception by examining how social networks affect correct voting through controlled experiments, demonstrating that social communication can serve as an effective shortcut for uninformed independents while potentially harming informed voters who incorporate biased partisan messages. However, this approach focuses on small group dynamics rather than large-scale network effects characterizing contemporary democratic communication. Most critically, no existing framework simultaneously models the beneficial and harmful aspects of networked communication within a unified theoretical structure.

We address these fundamental limitations through a novel agent-based modeling approach that explicitly incorporates both the epistemic benefits and systematic biases of networked political communication within a realistic framework of democratic choice. Our model makes three key theoretical innovations that advance understanding of democratic communication beyond existing scholarship. First, we formalize two mechanisms: honest sharing that filters noise and partisan bluffing that injects bias. Second, we demonstrate how network composition and architectural features—particularly the placement of independent actors in central positions—can serve as institutional solutions to the fundamental tension between these competing forces, providing concrete guidance for democratic design in networked environments. Third, we uncover a U-shaped pattern: accuracy collapses at moderate quality gaps because bias competes most effectively with truth in that range.

Our systematic computational analysis yields several striking findings that fundamentally challenge conventional wisdom about democratic communication while providing actionable insights for institutional reform. We demonstrate that extended deliberation can actually degrade collective judgment when partisan bluffing dominates honest information sharing, contradicting optimistic assumptions about the benefits of democratic discussion. We show that passionate partisans can sometimes outperform detached independents in identifying superior candidates, particularly when quality differences are small and partisan communication networks provide access to broader information sources. Our results show that network design outweighs voter traits for election quality, so reforms should target communication architecture instead of relying only on voter education. For digital democracy and platform policy, keeping independents in hub roles, timing deliberation carefully, and tracking extreme rhetoric may boost electoral accuracy better than standard voter-training efforts.

\section{Theory and Model}

\subsection{Theoretical Framework}

Democratic theory has long grappled with the conditions under which collective decision-making produces accurate outcomes. The Condorcet Jury Theorem establishes the foundational optimistic view: under assumptions of independent, unbiased information and sincere voting, majority rule generates increasingly accurate decisions as group size grows \citep{austen1996information}. This mathematical result suggests that diverse perspectives can systematically overcome individual cognitive limitations, forming the theoretical basis for democratic legitimacy through collective intelligence. However, real-world political communication violates these idealized assumptions in systematic ways. Citizens are embedded in social networks where information transmission occurs through endogenous trust relationships, partisan identities filter interpretation of evidence, and strategic considerations shape message construction and dissemination. Such departures can magnify accuracy or push beliefs off course, and the outcome hinges on which social mechanism dominates. We develop a framework in which two concurrent mechanisms fight for influence and jointly determine how accurate an electorate becomes. Understanding the relative strength of these mechanisms and the conditions determining which dominates represents a central challenge for democratic theory and institutional design.

At the core of our model are three most important components: Message, Trust, and Network. These elements interact to shape the process and outcome of collective decision-making.

\textbf{Message and Concurrent Mechanisms:}\\
\textbf{Mechanism 1: Noise Filtering.} The first mechanism captures the beneficial aspects of information aggregation emphasized in classical democratic theory. When voters receive independent, noisy signals about candidate quality and share them honestly through social networks, the law of large numbers ensures that idiosyncratic errors cancel out through repeated social interaction. If each voter observes $s_i=\theta+\epsilon_i$ with $\epsilon_i\sim\mathcal N(0,\sigma^{2})$, the sample mean $\bar s_n$ converges in probability to $\theta$ as $n$ grows. This convergence property holds regardless of the variance of individual errors, provided they remain independent and unbiased.

The noise filtering mechanism explains why diverse deliberation often improves collective judgment and why heterogeneous networks can outperform homogeneous ones in decision-making tasks. The key requirement is honest communication with independent errors across individuals. Noise filtering therefore predicts a monotone link between network size and belief precision, holding strategic distortion at zero.

\textbf{Mechanism 2: Partisan Bluffing.} The second mechanism reflects the strategic nature of political communication that formal models typically abstract away. Partisan actors derive utility not only from accurate candidate assessment but also from electoral victory for their preferred option. Because electoral victory carries private value, partisans gain by shading signals, which drags group beliefs away from ground truth. Let partisan $i$ supporting candidate $A$ receive private signal $s_i = \theta_A + \epsilon_i$ about candidate $A$'s quality. When communicating this information, the partisan faces a trade-off between honesty (promoting collective accuracy) and strategic bias (promoting electoral success). If the partisan values electoral victory with intensity $b_i > 0$, optimal communication involves systematically inflating the signal: $\text{transmitted message} = s_i + f(b_i)$ where $f(\cdot)$ is an increasing function of partisan intensity.

Unlike honest errors that cancel through aggregation, this strategic bias introduces systematic deviation from truth that compounds through network transmission. When multiple partisans interact, their individual biases reinforce rather than cancel, potentially leading to collective beliefs that diverge arbitrarily far from reality. The mathematical intuition is straightforward: if $n$ partisans each add bias $f(b_i)$ to their signals, the aggregate bias becomes $\sum_{i=1}^n f(b_i)$, which grows with network size rather than canceling out. This mechanism generates ``bluffing amplification"—a process by which strategic communication creates echo chambers that systematically distort collective beliefs, particularly when partisans cluster together in the network. The distortion effect grows super-linearly when biased agents cluster in the same neighborhood.

\textbf{Trust, Network Structure and Mechanism Dominance.} The relative strength of these competing mechanisms depends critically on trust structure, network structure and composition. Three factors prove particularly important:

\emph{Trust Structure:} The pattern of credibility assignment across different actor types determines how information flows through the network. If partisans systematically dismiss information from opponents while accepting biased signals from allies, partisan bluffing effects become self-reinforcing within homophilic clusters. This creates ``selective aggregation"—a process where voters aggregate information only from sources that confirm their biases, leading to polarized belief formation. The dynamic interplay between network structure and communication patterns reflects broader principles of network-behavior coevolution observed in social systems \citep{malik2016transitivity}.

\emph{Network Composition:} The proportion of honest versus strategic actors fundamentally shapes information flow patterns. When independents are plentiful, honest signals swamp bias and noise filtering wins. Conversely, networks with high proportions of strategic partisans amplify bias effects while providing limited noise reduction benefits, as the assumptions underlying beneficial aggregation are violated.

\emph{Network Centrality:} The partisan identity of highly connected actors can have disproportionate effects on system-wide outcomes due to their amplified influence over information diffusion. Independent actors in central positions serve as ``epistemic circuit breakers" that inject unbiased information into multiple partisan clusters simultaneously, while strategic partisans in central positions amplify bias throughout the network.

\textbf{Testable Predictions.} This theoretical framework generates precise, testable predictions:

\begin{enumerate}
\item \textbf{Noise filtering dominates when:} networks contain high proportions of independent actors, trust patterns enable cross-cutting communication, and honest actors occupy central positions.

\item \textbf{Partisan bluffing dominates when:} networks are populated primarily by strategic partisans, trust patterns create isolated homophilic clusters, and strategic actors control key network positions.

\item \textbf{Interaction effects:} The transition between regimes depends on candidate quality differences, distribution of partisan intensities, and network topology.
\end{enumerate}

\subsection{Agent-Based Model}

To rigorously test these theoretical predictions, we develop a formal agent-based model that explicitly incorporates both mechanisms within a realistic network environment. Our modeling approach balances theoretical precision with empirical realism, enabling systematic isolation of causal effects while capturing essential features of democratic communication.

\subsubsection{Model Setup and Primitives}

\textbf{Population Structure.} We model $n = 1000$ voters choosing between candidates $A$ and $B$. The proportion of partisans of each party, and independent voters, $\tau_i \in {A, B, I}$, is drawn from a discrete uniform distribution, though later experiments relax this share to test robustness. This tripartite division captures the empirical reality that democratic electorates contain both committed partisans and less aligned citizens, while equal proportions provide a neutral baseline for examining network effects.

\textbf{Network Architecture.} Voters are connected through an undirected social network $G = (V, E)$ generated using the Barabási-Albert preferential attachment model with average degree $\langle k \rangle = 6$, producing scale-free degree distributions that match empirical observations of real social networks. The preferential attachment process creates networks with realistic heterogeneity in connectivity while avoiding extreme clustering or randomness that might artificially bias results.

\subsubsection{Candidate Quality and Voter Utilities}

Each candidate $X \in \{A, B\}$ provides two distinct benefit types:

\emph{Global Benefits} ($GB_X$): Policy outcomes valued equally by all voters, drawn from Uniform[20, 80]. These represent the ``common good" component of electoral choice.

\emph{Partisan Benefits} ($PB_i$): Group-specific advantages for partisan voters from Uniform[1, 10]; independents receive $PB_i = 0$.

Voter $i$'s utility if candidate $A$ wins:
\begin{align}
U_i(A \text{ wins}) = \begin{cases}
GB_A + PB_i & \text{if } \tau_i = A \\
GB_A - PB_i & \text{if } \tau_i = B \\
GB_A & \text{if } \tau_i = I
\end{cases}
\end{align}
This specification formalizes the tension between shared interests in good governance and group-specific electoral preferences that drives strategic communication incentives.

\subsubsection{Information Structure and Belief Formation}

At $t = 0$, voters receive $k_i \sim \text{Uniform}\{1,2,3,4\}$ private signals about both candidates:
\[
s_{i,\ell}^X = GB_X + \epsilon_{i,\ell}, \quad \epsilon_{i,\ell} \sim \mathcal{N}(0, \sigma^2 = 1600)
\]
Voters form initial beliefs by averaging private signals: $\widehat{GB}_{i,0}^X = \frac{1}{k_i} \sum_{\ell=1}^{k_i} s_{i,\ell}^X$. The high noise variance creates substantial uncertainty that makes social learning potentially valuable.

\subsubsection{Communication and Strategic Behavior}

The core innovation of our model lies in the communication process, which formally captures both honest information sharing and strategic bias. In each round $t \geq 1$, every voter $i$ simultaneously broadcasts signals about both candidates to all network neighbors. Signal construction depends on voter type:
\begin{align}
\text{Partisan A: } \quad \tilde{s}_{i,t}^A &= \widehat{GB}_{i,t}^A + PB_i, \quad \tilde{s}_{i,t}^B = \widehat{GB}_{i,t}^B - PB_i \\
\text{Partisan B: } \quad \tilde{s}_{i,t}^A &= \widehat{GB}_{i,t}^A - PB_i, \quad \tilde{s}_{i,t}^B = \widehat{GB}_{i,t}^B + PB_i \\
\text{Independent: } \quad \tilde{s}_{i,t}^A &= \widehat{GB}_{i,t}^A, \quad \tilde{s}_{i,t}^B = \widehat{GB}_{i,t}^B
\end{align}
This formalization is consistent with experimental evidence from \cite{ryan2011social}, but our model explicitly captures how individual-level strategic distortions aggregate through repeated network transmission to create system-wide patterns of collective delusion.

Voters update beliefs using trust-weighted averaging:
\[
\widehat{GB}_{i,t+1}^X = \frac{\widehat{GB}_{i,t}^X + \delta \sum_{j \in N_i} w_{\tau_i,\tau_j} \cdot \tilde{s}_{j,t}^X}{1 + \delta \sum_{j \in N_i} w_{\tau_i,\tau_j}}
\]
where $\delta = 0.1$ controls adaptation rate and the trust matrix reflects realistic credibility patterns:

\begin{align*}
W = 
\begin{array}{c|ccc}
  & \multicolumn{3}{c}{\textbf{Sender}} \\
\textbf{Receiver} & \text{A} & \text{B} & \text{I} \\
\hline
\text{A} & 1.0 & 0.0 & 0.5 \\
\text{B} & 0.0 & 1.0 & 0.5 \\
\text{I} & 0.0 & 0.0 & 1.0 \\
\end{array}
\end{align*}

These weights encode four stylized facts: co-partisans gain full trust, opponents gain none, and independents gain partial trust from partisans. For independents, trust is mutual among themselves, but they do not trust partisans. We also assume that each node's partisanship is common knowledge; that is, both sender and receiver know the partisanship affiliation of all others and apply the appropriate trust parameter when updating their beliefs.
\subsubsection{Voting Behavior and Performance Measurement}

After $T = 500$ communication rounds, voters cast ballots based on final beliefs and utility calculations. We define correct voting as the choice each voter would make under perfect information, providing an objective standard accounting for both shared interests and legitimate partisan considerations. Our primary outcome is the system-wide correct voting rate: $\overline{CV} = \frac{1}{n}\sum_{i=1}^n CV_i \in [0,1]$. This measure allows us to assess how different network configurations, communication patterns, and candidate characteristics affect collective democratic performance. Values near 1 indicate high democratic competence, while values near 0 suggest systematic collective error.

\subsubsection{Model Limitation}
We formalize individuals’ strategic behaviors within Ryan’s framework \cite{ryan2011social} using a simulation approach. In contrast to classical models, which often abstract away explicit strategic considerations, we impose a fixed set of behavioral rules. Specifically, we assume that partisans consistently send the same signals and maintain constant trust parameters over time. This simplification is justified by our primary goal: exploring how social network characteristics interact with predefined, simple strategies.

Nevertheless, we acknowledge that strategic behaviors in reality are likely endogenous, adapting dynamically to changes in the network environment and other agents’ actions. Addressing such complexity would typically involve equilibrium solution concepts, such as Nash or Bayesian Nash equilibrium. As such strategic adaptiveness is beyond the intended scope of this paper, we clearly note our fixed-strategy assumption as a limitation and suggest that future research extend our analysis to incorporate endogenous strategy formation.

\section{Results}

Our agent-based experimental analysis examines when communication networks enhance versus undermine democratic decision-making through comprehensive simulations. We present results from over 10,000 simulation runs across systematically varied parameter configurations, each involving 1,000 voters communicating through 500 rounds before casting ballots. Each run settles into a steady state long before round 500, so every reported pattern comes from equilibrium, not noise. This extensive computational analysis reveals striking patterns that challenge conventional wisdom about information aggregation while providing precise conditions under which each theoretical mechanism dominates.

\subsection{The Fundamental Tension: When Communication Networks Turn Against Democracy}

\begin{figure}[!ht]
 \centering
 \caption{The Dual Nature of Political Communication: A Representative Simulation}
 \label{fig:bench}
 
 \begin{subfigure}{0.49\textwidth}
     \includegraphics[width=0.8\linewidth]{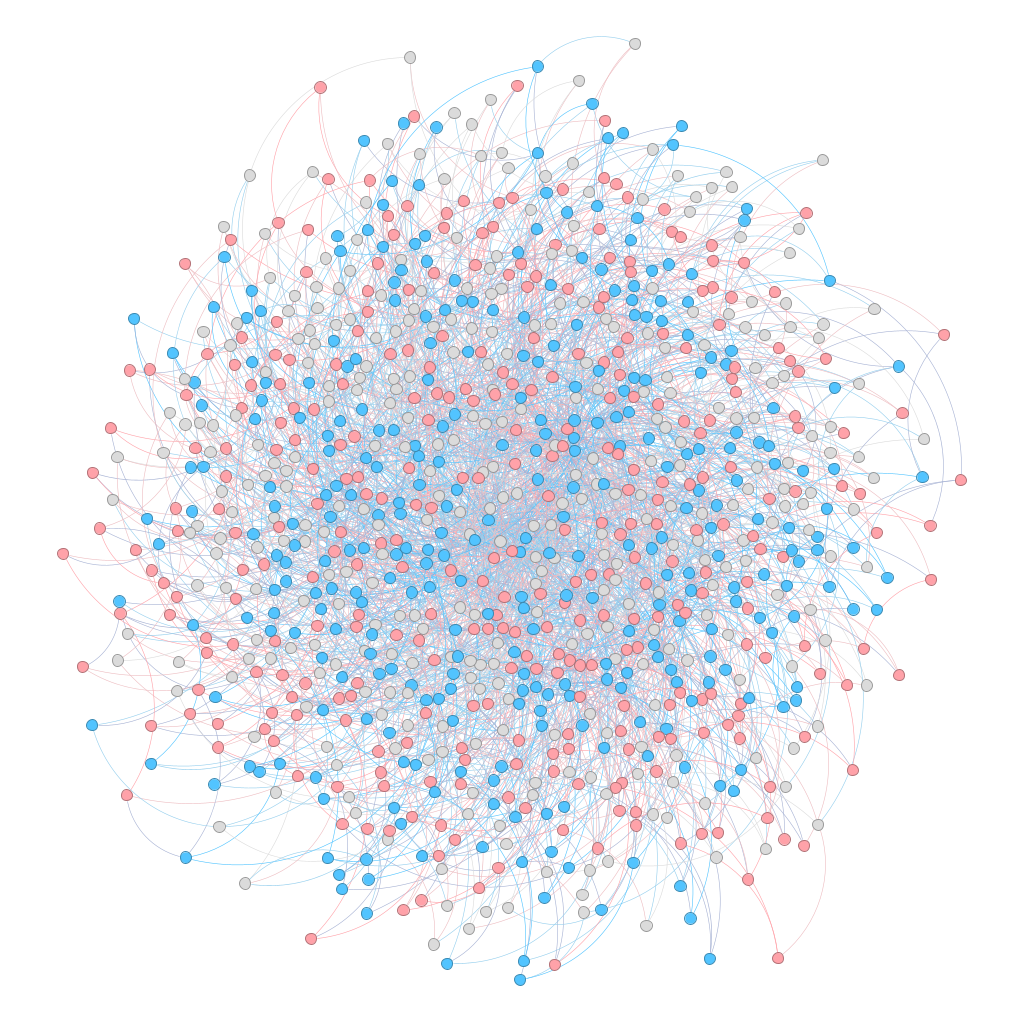}
     \caption{Network Structure and Partisan Assignment}
     \label{fig:bench_nt}
 \end{subfigure}
 \hfill
 \begin{subfigure}{0.49\textwidth}
     \includegraphics[width=\linewidth]{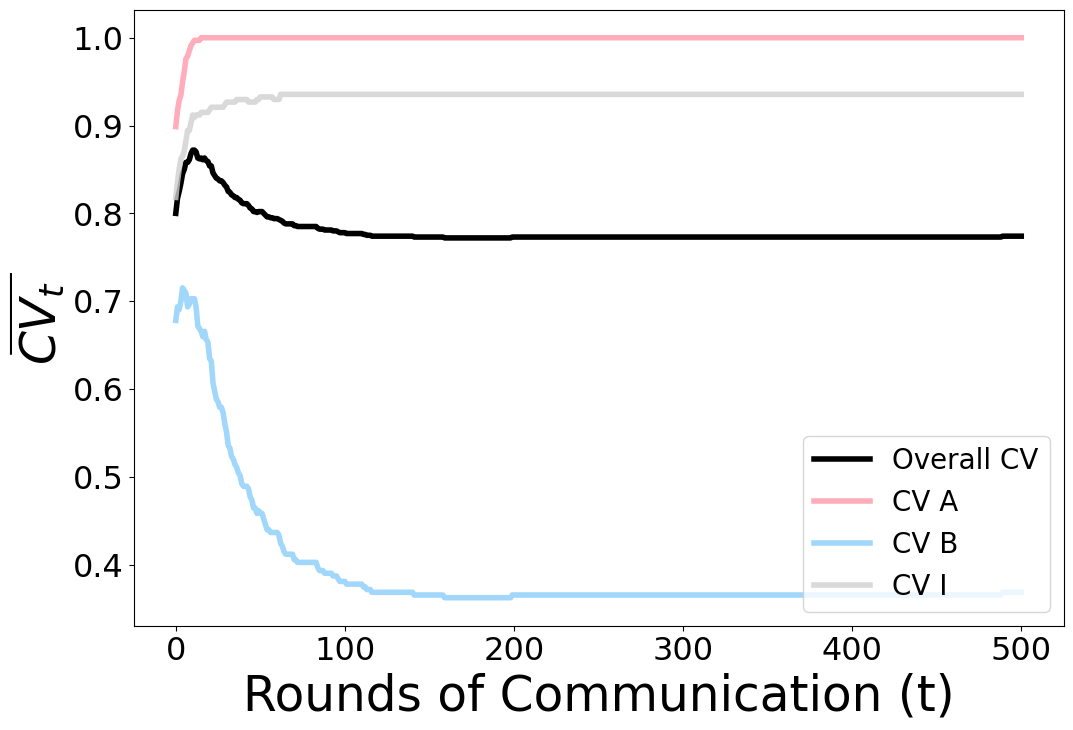}
     \caption{Evolution of Correct Voting Rates}
     \label{fig:bench_cv}
 \end{subfigure}
 
 \bigskip
 
 \begin{subfigure}{0.49\textwidth}
     \includegraphics[width=\linewidth]{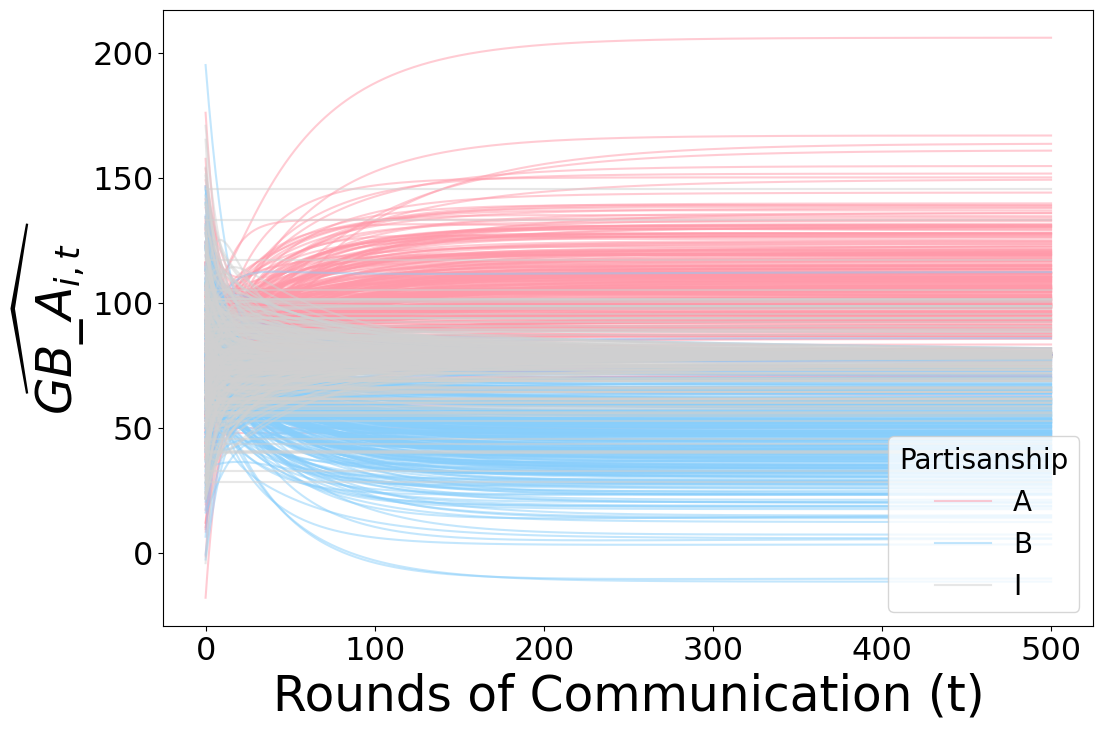}
     \caption{Belief Evolution: Candidate A Quality}
     \label{fig:bench_GBAh}
 \end{subfigure}
 \hfill
 \begin{subfigure}{0.49\textwidth}
     \includegraphics[width=\linewidth]{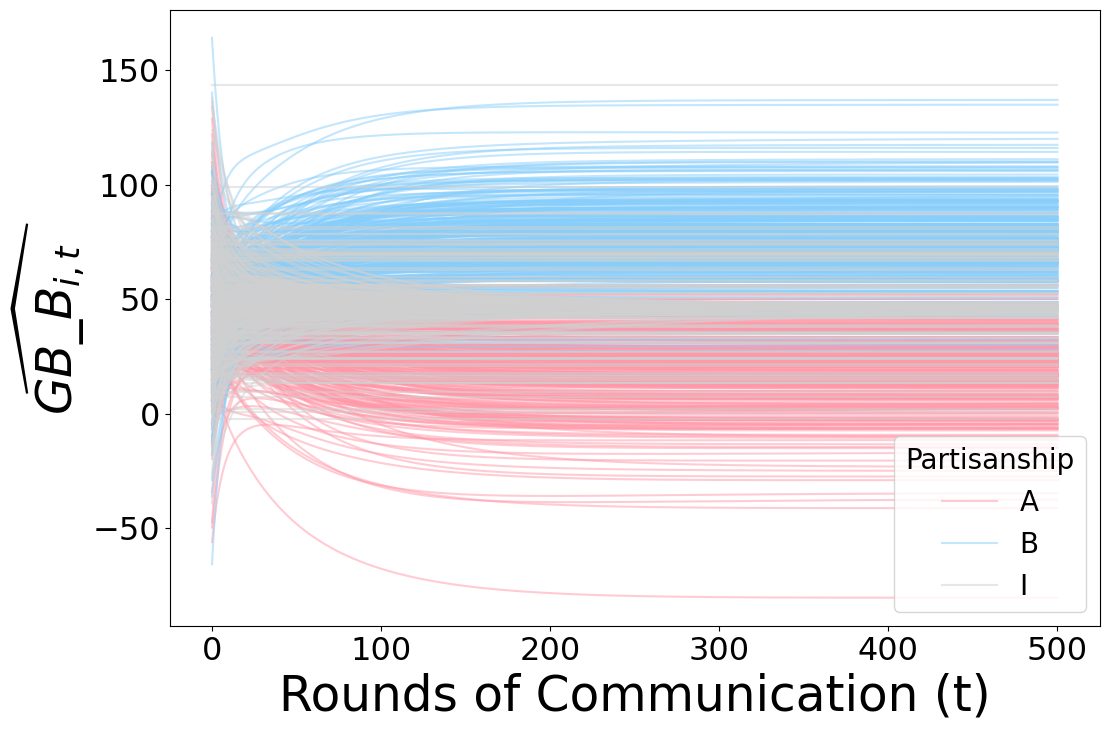}
     \caption{Belief Evolution: Candidate B Quality}
     \label{fig:bench_GBBh}
 \end{subfigure}
\end{figure}

To establish core dynamics of our theoretical framework, we begin with detailed analysis of a representative simulation exemplifying the fundamental tension between beneficial information aggregation and harmful partisan bias. Figure \ref{fig:bench} presents results from a simulation where true global benefits are $GB_A = 77.9$ and $GB_B = 46.4$, making candidate A objectively superior by 31.5 units.

Figure \ref{fig:bench}(a) shows network visualization with partisan A as red nodes, partisan B as blue nodes, and independents as gray nodes. Figure \ref{fig:bench}(b) reveals the profound divergence in how communication affects different voter types, providing striking evidence for our dual-mechanism framework. Initially, all voter groups exhibit rising correct voting rates as social learning filters noise from private information signals, reflecting beneficial effects of information aggregation where independent errors cancel through communication as predicted by classical models of collective intelligence. Improvement is particularly pronounced for partisans during the first 20 rounds, as they access broader information networks including both co-partisans and independents, providing larger effective sample sizes for noise reduction. After round 20, results split. Independents keep improving, but partisan-B voters slip to about 0.20 correct and never recover. The pattern shows that strategic bias can drag a group below the accuracy of isolated individuals, creating a stable but mistaken equilibrium.

The underlying mechanisms driving this divergence reflect two fundamentally different approaches to political communication operating simultaneously within the same network structure. Independent voters illustrate how honest sharing cleans the signal and lifts accuracy. When independents communicate signals honestly, the Law of Large Numbers operates to filter random errors while preserving true candidate quality signals. This manifests in Figure \ref{fig:bench}(b) as monotonic improvement from 0.80 to nearly 0.90 by round 500, demonstrating collective wisdom emerging from individual ignorance as each round of information sharing reduces idiosyncratic noise influence while amplifying the common signal reflecting genuine candidate differences. The mechanism succeeds because independent voters have aligned incentives, benefiting only from accurate candidate assessment and facing no strategic reason to distort shared information, creating a virtuous cycle where better information leads to better decisions.

In contrast, partisan voters face incentives that transform communication into systematic bias. Partisan B supporters systematically overestimate their candidate's benefits through strategic ``bluffing"—inflating signals about their preferred candidate while deflating opponent assessments. This creates ``bluffing amplification" where bias compounds through network transmission rather than canceling like random noise. This strategic distortion creates what we term ``bluffing amplification," where bias magnitude depends critically on local network composition surrounding each voter: when a partisan B voter is surrounded primarily by independents, honest signals from independents provide substantial corrective influence limiting bias in updated beliefs, but when the voter is embedded in a cluster of co-partisans, biased signals from co-partisans dominate the information environment, leading to severely distorted beliefs about candidate quality. Under extreme conditions where stabilizing mechanisms are removed entirely—such as networks with very low connectivity, purely partisan populations, or configurations where independents lose their credibility—this bluffing process can lead to unbounded belief divergence where partisan assessments spiral completely away from reality, as demonstrated in the comprehensive analysis presented in Appendix Figure \ref{fig:unbounded_belief}.

Figures \ref{fig:bench}(c) and \ref{fig:bench}(d) illuminate how competing forces shape collective beliefs. Independent voters' beliefs converge toward true values (A→77.9, B→46.4), while partisan B networks develop distorted beliefs (B→90, A→40) that persist despite continued communication. The stabilization rather than unlimited divergence reveals independents' role as ``epistemic circuit breakers," providing corrective information that bounds collective delusion while maintaining partial trust from partisans. Since independents only trust other independents and remain immune to partisan bluffing dynamics, they maintain accurate beliefs throughout the communication process, and because partisans assign positive weight to independent signals, these honest voices provide gravitational pull toward truth preventing partisan beliefs from spiraling completely out of control.

\subsection{The Paradox of Moderate Competition: When Democratic Choice Becomes Most Difficult}

One of our most surprising and policy-relevant findings concerns the relationship between objective candidate quality differences and democratic performance. Common intuition says big quality gaps should make correct voting easier because the better candidate is obvious. Our systematic analysis across 2,000 simulations reveals an unexpected relationship between candidate quality differences and democratic performance that challenges conventional wisdom.

\begin{figure}[H]
 \centering
 \caption{The Surprising Challenge of Moderate Competition}
 \label{fig:GB}
 
 \begin{subfigure}[t]{0.48\textwidth}
     \includegraphics[width=\linewidth]{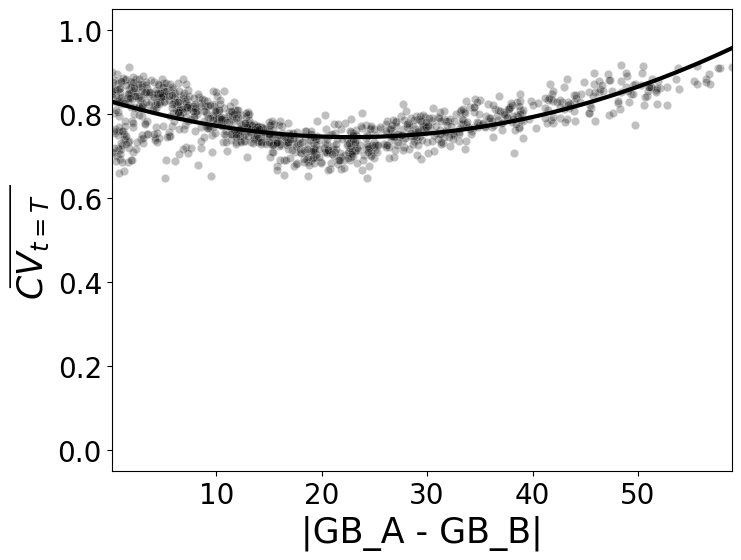}
     \caption{Aggregate Correct Voting Performance}
     \label{fig:GB_overall}
 \end{subfigure}
 \hfill
 \begin{subfigure}[t]{0.48\textwidth}
     \includegraphics[width=\linewidth]{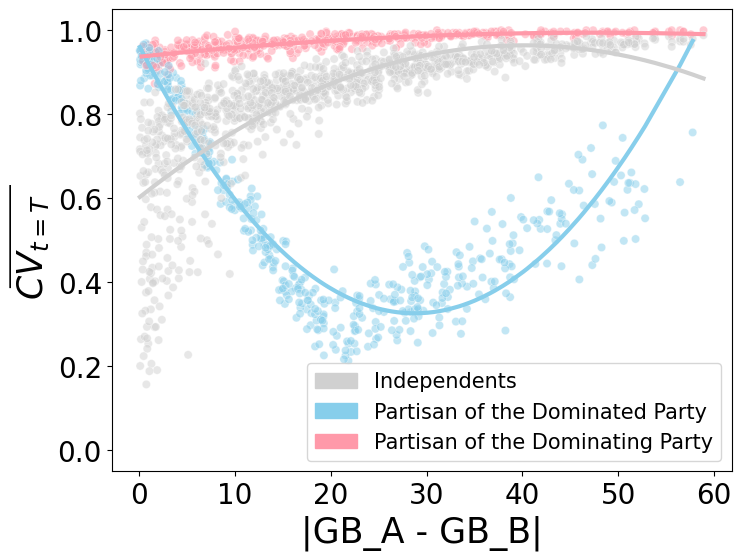}
     \caption{Performance by Voter Type}
     \label{fig:GB_partisan}
 \end{subfigure}
\end{figure}

Figure \ref{fig:GB} presents our core finding: a robust U-shaped relationship between candidate quality differences and collective voting accuracy that holds across multiple specifications and robustness checks. The aggregate pattern reveals a pronounced U-shaped curve with minimum occurring when global benefit differences reach moderate levels around 20-30 units. Performance deteriorates from 0.60 for minimal differences to 0.50 at moderate levels, recovering to 0.80 for extreme differences above 60 units. This contradicts standard models and reveals a fundamental paradox: competitive elections with meaningful but not overwhelming quality differences may be most vulnerable to network-induced errors—precisely the circumstances where accurate collective choice should matter most for democratic governance.

Figure \ref{fig:GB}(b) reveals the mechanisms generating this pattern, disaggregating results by voter type to illuminate why different regions of the candidate quality distribution produce such different democratic outcomes. When differences are small (left region, $|GB_A - GB_B| < 15$), independents struggle to distinguish between candidates because the true signal is weak relative to the noise variance ($\sigma^2 = 1600$), and in this low-signal regime, partisan heuristics prove beneficial as partisans outperform independents through broader information networks and useful tie-breaking. The problematic regime emerges at moderate levels where partisan bias ($\mathbb{E}[PB_i] = 5.5$) becomes comparable to true differences—creating effective bias of 11 units that overcomes genuine signals through network amplification. Recovery occurs when overwhelming evidence cannot be masked by systematic bias, as truth overwhelms bias when signal strength greatly exceeds strategic distortion magnitude, leading all voter types to converge on correct choices despite different communication incentives.

\subsection{The Distribution of Partisan Benefits: When Extreme Preferences Destabilize Democracy}

While our baseline analysis employs uniform distributions for partisan benefits to ensure analytical tractability, real-world partisan preferences likely follow more realistic distributional patterns that can significantly affect democratic outcomes. We examine how partisan benefit distributions affect democratic outcomes using Gamma(1, $\beta$) distributions that realistically represent voter preferences.

\begin{figure}[h] 
  \centering 
  \caption{Correct Voting and the Distribution of Partisan Benefits}
  \label{fig:pb}

  \begin{subfigure}{0.55\textwidth}
      \includegraphics[width= \linewidth]{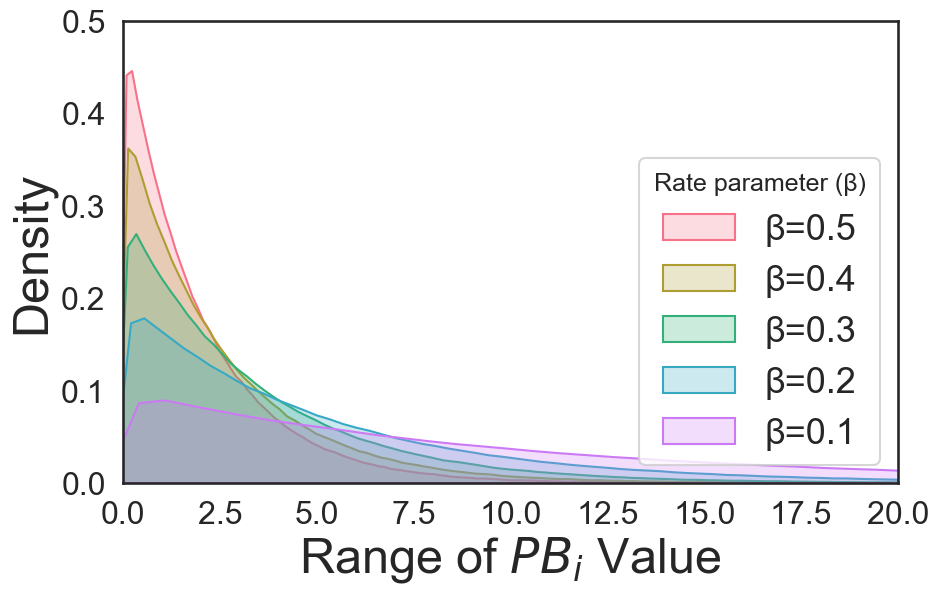}
      \caption{Distribution of Partisan Benefits}
      \label{subfig:beta_dist}
  \end{subfigure}
  
  \vspace{1cm}
  
  \begin{subfigure}{0.49\textwidth}
      \includegraphics[width= \linewidth]{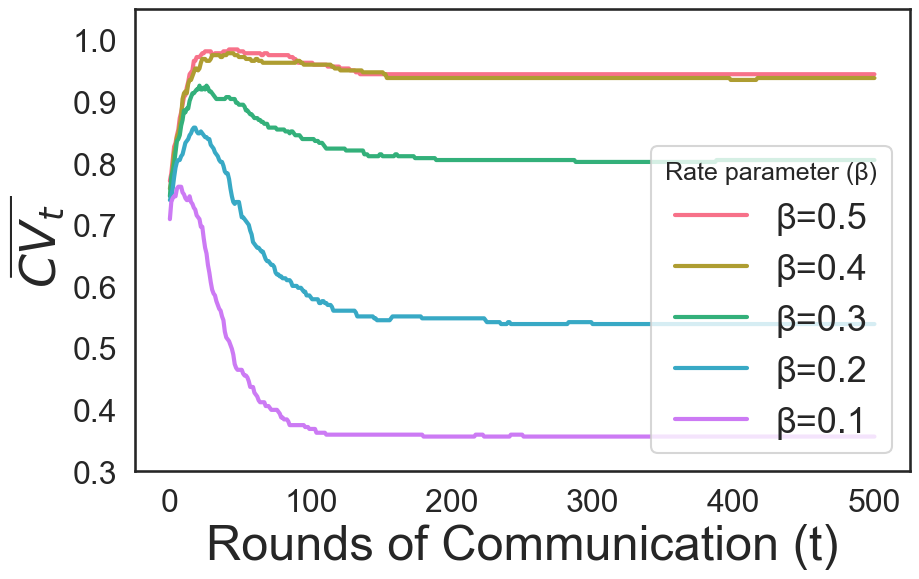} 
      \caption{Correct Voting: Supporters of Inferior Candidate} 
      \label{subfig:PB_dominated}
  \end{subfigure}
  \hfill 
  \begin{subfigure}{0.49\textwidth}
      \includegraphics[width= \linewidth]{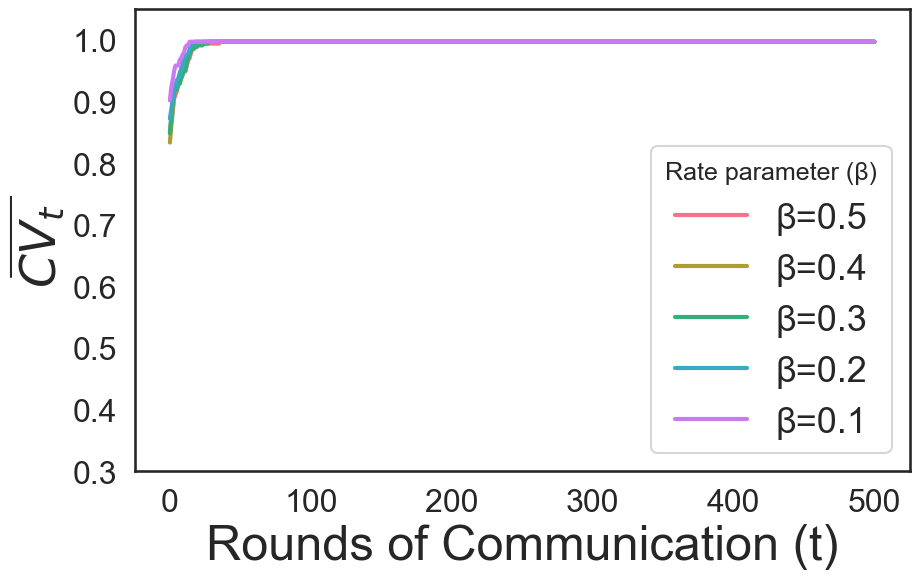}
      \caption{Correct Voting: Supporters of Superior Candidate} 
      \label{subfig:PB_dominating}
  \end{subfigure}
\end{figure}

The Gamma distribution reflects Median Voter Theorem expectations: most voters experience small partisan benefits while a small fraction holds extreme preferences \citep{caplin1991aggregation}. Parameter $\beta$ controls distribution spread, with smaller values creating more extreme outliers. This choice reflects theoretical expectations from the Median Voter Theorem: in competitive two-party systems, both parties position themselves close to the median voter to maximize electoral support, resulting in most voters experiencing relatively small partisan benefits while a small fraction holds extreme preferences.

Using our benchmark scenario ($GB_A = 77.9$, $GB_B = 46.4$), Figures \ref{fig:pb}(b) and \ref{fig:pb}(c) reveal striking patterns. For supporters of the inferior candidate, correct voting rates collapse dramatically as $\beta$ decreases below 0.3, falling from nearly perfect performance to systematic collective delusion. This reflects extreme partisans' amplified influence through aggressive bluffing that contaminates belief formation throughout network neighborhoods. Individual-level mechanisms underlying this collective deterioration become clear through detailed examination of belief trajectories across different distributional configurations, as illustrated in Appendix Figure \ref{fig:PB_belief}, demonstrating how extreme partisan preferences function as ``bias super-spreaders" contaminating collective belief formation far beyond their numerical representation through cascading effects in network communication. Supporters of the superior candidate maintain stable performance, benefiting from alignment between partisan biases and objective reality. 

The asymmetric effects reveal democratic vulnerability: supporters of inferior candidates prove systematically more susceptible to bias amplification, creating conditions where small numbers of extreme actors generate system-wide failures through cascading bias amplification. When extreme partisan B supporters inflate their candidate's signals by large amounts while deflating opponent signals correspondingly, their biased communications reach moderate partisans who lack extreme preferences necessary to discount these distorted messages appropriately. This creates cascading bias amplification where extreme preferences at the distributional tail contaminate belief formation among moderate voters, leading entire partisan communities toward collective delusion about candidate quality. The comprehensive analysis in Appendix Figure \ref{fig:PB_3D} reveals how distributional effects interact with candidate quality differences to create performance cliffs.

\subsection{Network Composition as Democratic Infrastructure: The Stabilizing Power of Independent Voices}

\begin{figure}[!ht]
 \centering
 \caption{Independent Voters as Democratic Stabilizers}
 \label{fig:size}
 
 \begin{subfigure}{0.49\textwidth}
     \includegraphics[width=\linewidth]{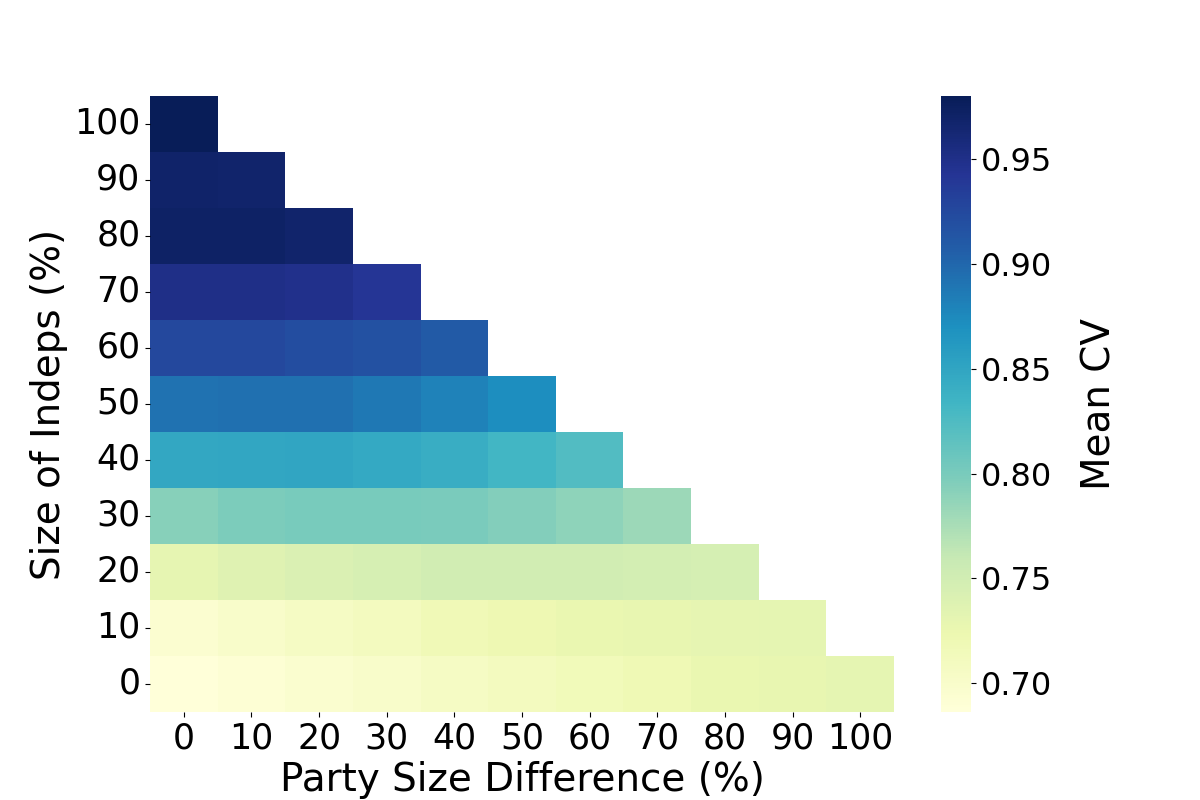}
     \caption{Mean Correct Voting Performance}
     \label{subfig:size_mean}
 \end{subfigure}
 \hfill
 \begin{subfigure}{0.49\textwidth}
     \includegraphics[width=\linewidth]{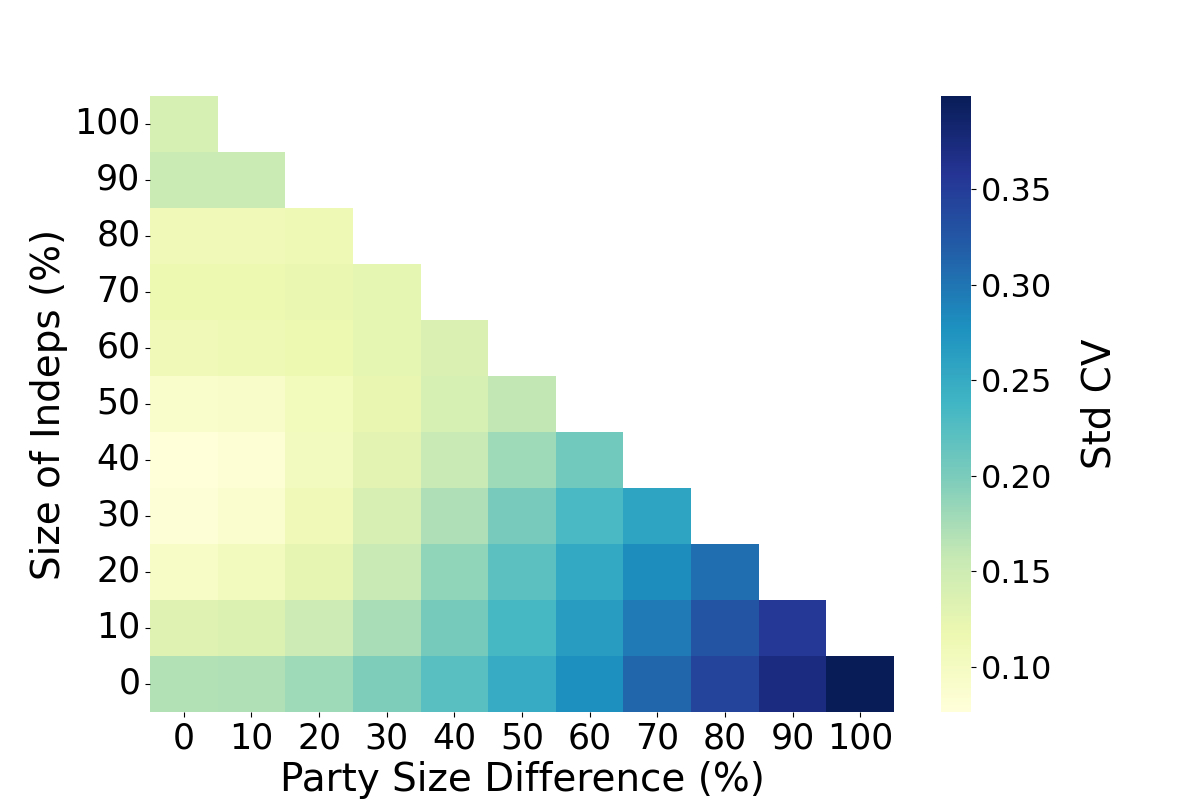}
     \caption{Performance Stability Across Elections}
     \label{subfig:size_sd}
 \end{subfigure}
\end{figure}

Our analysis reveals that the proportion of independent voters in society functions as critical democratic infrastructure, with effects on both the level and stability of collective decision-making that exceed those of individual voter characteristics. Figure \ref{fig:size} presents results from simulations varying independent proportions from one-sixth to one-half while holding other parameters constant.

We find a simple rule: more independents raise mean accuracy and reduce volatility. As independent share rises from one-sixth to one-half, mean correct voting increases from 0.55 to 0.75 while standard deviation decreases from 0.12 to 0.06.

This reflects two mechanisms. First, independents provide ``epistemic gravity" toward truth through unbiased information. Unlike partisans who strategically distort signals based on electoral preferences, independents report genuine beliefs about candidate quality without systematic bias, creating consistent pull toward accuracy benefiting all network participants. Second, they function as ``epistemic circuit breakers" preventing complete partisan isolation. The mathematical intuition: when independent proportion is $p_I$, expected bias reduces by factor $(1-p_I)$, creating multiplicative accuracy effects. Because partisans assign partial trust to independent voices, independents ensure that even heavily partisan network clusters receive corrective information moderating the most extreme biases.

These findings challenge assumptions about political engagement, suggesting societies benefit from maintaining significant populations prioritizing accuracy over group loyalty. The implications prove urgent given declining independent proportions in established democracies as partisan sorting intensifies. While passionate partisan involvement is often celebrated as a hallmark of democratic health, our analysis suggests societies benefit substantially from maintaining significant populations of less partisan citizens who prioritize accuracy over group loyalty in political communication.

\subsection{Network Architecture and Democratic Performance: Who Controls the Information Flow}

\begin{figure}[H]
 \centering
 \caption{Network Centrality and the Architecture of Democratic Communication}
 \label{fig:part_cent}

 \begin{subfigure}{0.49\textwidth}
     \includegraphics[width=\linewidth]{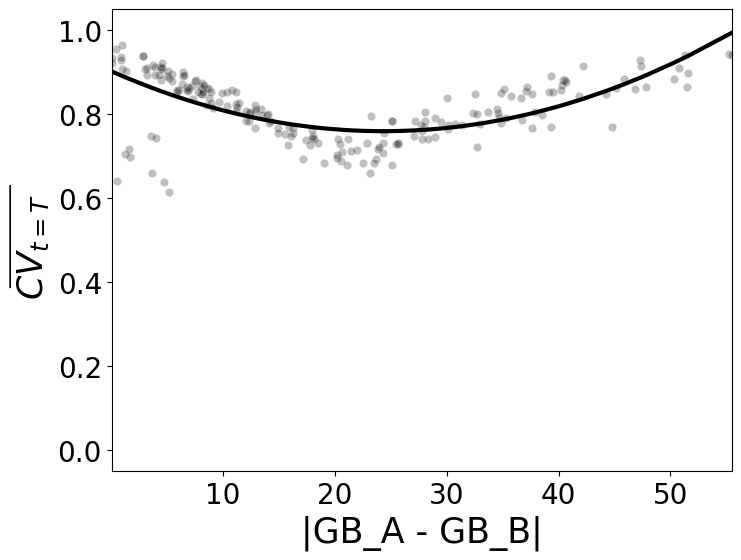}
     \caption{Random Assignment (Baseline)}
 \end{subfigure}
 \hfill
 \begin{subfigure}{0.49\textwidth}
     \includegraphics[width=\linewidth]{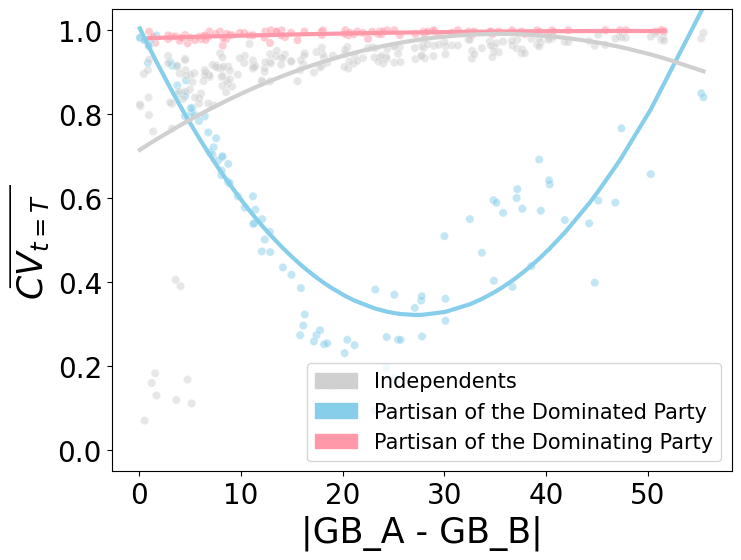}
     \caption{Random Assignment by Group}
 \end{subfigure}
 
 \vspace{1em}
 
 \begin{subfigure}{0.49\textwidth}
     \includegraphics[width=\linewidth]{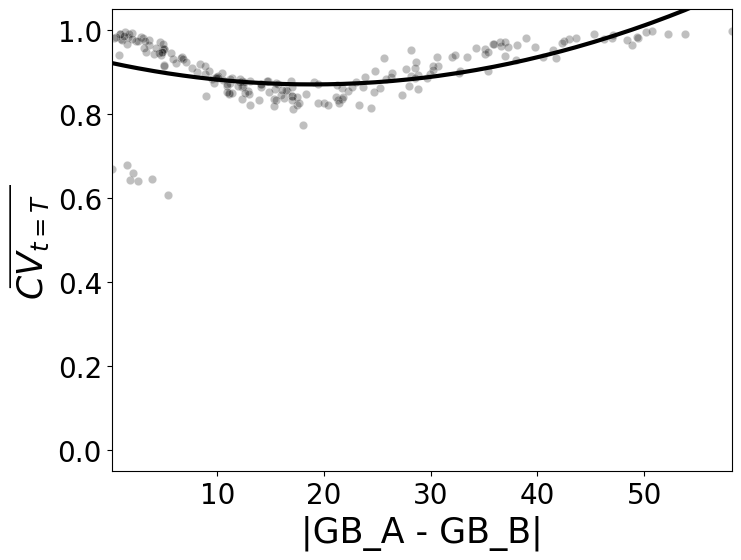}
     \caption{Independents in Central Positions}
 \end{subfigure}
 \hfill
 \begin{subfigure}{0.49\textwidth}
     \includegraphics[width=\linewidth]{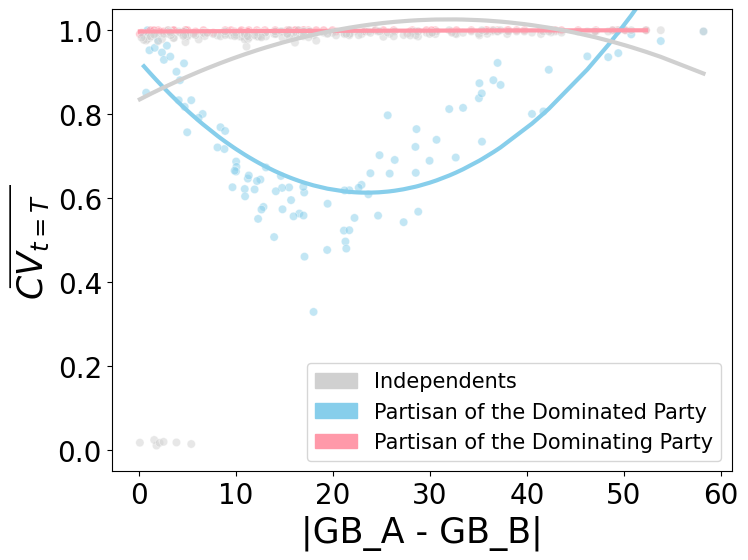}
     \caption{Independents Central, by Group}
 \end{subfigure}
 
 \vspace{1em}
 
 \begin{subfigure}{0.49\textwidth}
     \includegraphics[width=\linewidth]{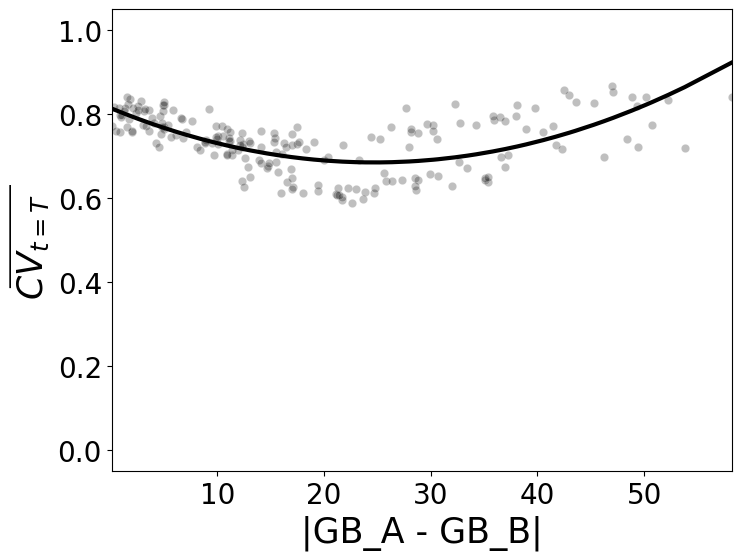}
     \caption{Partisans in Central Positions}
 \end{subfigure}
 \hfill
 \begin{subfigure}{0.49\textwidth}
     \includegraphics[width=\linewidth]{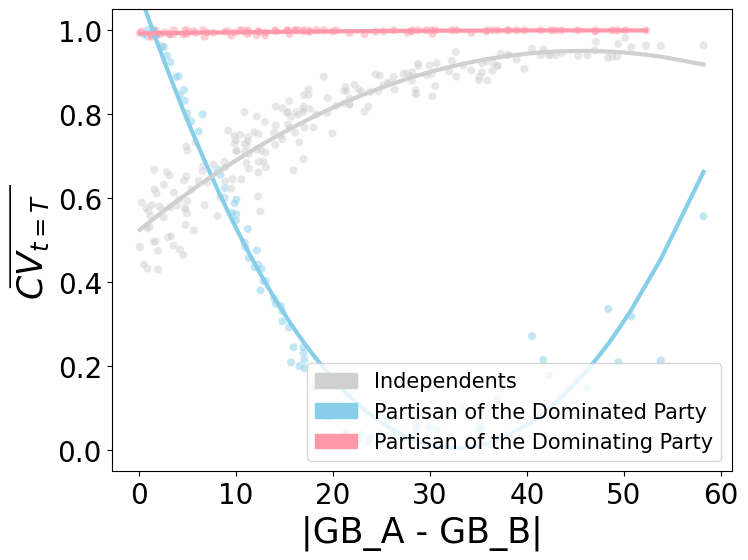}
     \caption{Partisans Central, by Group}
 \end{subfigure}
\end{figure}

Perhaps our most striking and policy-relevant finding concerns the disproportionate importance of network centrality in determining system-wide democratic outcomes. The partisan identity of highly connected actors can dramatically affect collective decision-making quality, often overwhelming effects of aggregate network composition or individual voter characteristics. Figure \ref{fig:part_cent} compares performance across three architectures varying partisan composition of central positions.

The top row establishes our baseline scenario where partisan identities are randomly assigned to network positions without regard to centrality. Under this configuration, we observe the characteristic U-shaped relationship between candidate quality differences and democratic performance that emerged in our earlier analysis. Positioning independents centrally (middle row) produces dramatic improvements across virtually all quality difference levels, rising from baseline 0.75-0.95 to consistently elevated 0.80-1.0 range. Central independents inject unbiased information into multiple partisan clusters simultaneously, moderating bluffing dynamics through consistent corrective signals. The mechanism underlying this improvement reveals fundamental insights about information diffusion in political networks: highly connected independent nodes function as ``epistemic circuit breakers" injecting unbiased information into multiple partisan clusters simultaneously.

Concentrating partisans centrally (bottom row) produces substantial deterioration, declining to 0.60-0.80 range. Highly connected partisan nodes amplify strategic bias throughout the system, creating ``systemic misinformation" where biased central actors contaminate information flow across the entire democratic system. These ``super-spreaders" of partisan bias reach voters who would otherwise receive more balanced information, effectively expanding echo chamber reach beyond their natural boundaries and creating what we term ``systemic misinformation"—false beliefs persisting not because of isolated echo chambers but because biased central actors contaminate information flow across the entire democratic system.

These findings reveal a crucial dimension of democratic design: the identity of influential communicators may matter more than aggregate numbers in determining collective decision-making quality. A small proportion of independent voices strategically positioned in central network locations can stabilize an entire democratic system, while a similarly small proportion of strategic actors in those same positions can destabilize it fundamentally.

\subsection{Robustness and Boundary Conditions}

Extensive robustness analyses across alternative parameter specifications, network structures, and model assumptions confirm our core results represent fundamental features rather than modeling artifacts. Our core results prove remarkably stable across these variations, suggesting that the dual mechanisms we identify represent fundamental features of political communication rather than artifacts of specific modeling choices.

Varying trust matrix parameters within plausible bounds produces consistent qualitative patterns while affecting effect magnitudes. Alternative network structures—including Erdős-Rényi random graphs, small-world networks, and empirically calibrated social media networks—produce qualitatively similar results with quantitative differences. The U-shaped relationship emerges consistently across all tested topologies, though the precise location of the minimum varies with structural parameters such as clustering coefficients and degree distributions. Scale-free networks with high clustering amplify both beneficial aggregation among independents and harmful bias amplification among partisans, while more homogeneous structures moderate both effects but preserve the fundamental patterns we identify.

Sensitivity checks show our results hold as long as the model allows both noise filtering and bluffing to operate. Extreme values can eliminate network effects: very low connectivity reduces to individual decision-making, minimal partisan benefits eliminate strategic incentives, while overwhelming noise drowns out signals. These checks confirm our framework captures general features extending beyond specific institutional contexts, providing confidence in broader applicability of our insights to real-world democratic systems where voters face genuine uncertainty about candidate quality while maintaining distinct preferences over electoral outcomes.

\section{Discussion and Implications}

Our analysis fundamentally reconceptualizes the relationship between communication networks and democratic performance by revealing how the same structural features that enable beneficial information aggregation can simultaneously generate systematic bias that undermines collective decision-making. Our two-mechanism framework clarifies why networks sometimes lift and sometimes sink electoral competence. By demonstrating that strategic partisan bluffing can overpower honest information sharing precisely when accurate choice matters most—during moderately competitive elections—we challenge core assumptions underlying classical theories of democratic deliberation and reveal previously unrecognized vulnerabilities in networked democratic systems. The vision that more connectivity always helps democracy is too rosy; common network settings can push citizens toward systematic error.

The theoretical implications extend far beyond the specific mechanisms we model to encompass fundamental questions about the microfoundations of democratic legitimacy. Unlike existing accounts of political polarization that rely on psychological biases, motivated reasoning, or selective media exposure, we demonstrate how purely strategic communication incentives can generate persistent collective delusion even when all actors receive identical initial information and possess equal cognitive capabilities. This departure from psychological explanations aligns with growing evidence that partisan polarization operates through social identity mechanisms \citep{iyengar2015fear}, that even corrections of political misperceptions can backfire when they threaten partisan identities \citep{nyhan2010corrections}, and that political learning often involves opinion change rather than simple priming effects \citep{lenz2009learning}. Because the failure is structural, fixes that target only voter skills, such as civic courses or fact-checks, will not be enough. While \cite{ryan2011social} provided experimental evidence that social communication can harm democratic decision-making when voters receive biased information from opposing partisans, our theoretical framework reveals the underlying network mechanisms that determine when these harmful effects systematically dominate beneficial information aggregation across entire democratic systems. These findings complement research showing how network disagreement affects political participation and how institutional design can shape the quality of democratic decision-making across different contexts. We show that who sits at the network core and how trust moves between blocs are pivotal for election quality.

Perhaps our most counterintuitive finding concerns the U-shaped relationship between candidate quality differences and democratic accuracy, which reveals that competitive elections may be systematically more vulnerable to network-induced errors than either very close races or overwhelming landslides. The finding overturns the idea that more competition always helps; mid-level contests invite bluffing that swamps truth. The mathematical intuition is straightforward but profound: when true candidate differences are comparable to the magnitude of strategic bias that partisans introduce into their communications, repeated network transmission can systematically distort collective beliefs away from objective reality. This finding has important implications for understanding democratic performance across different institutional contexts \citep{lau2014correct} and suggests that the relationship between electoral competition and democratic quality may be more complex than traditionally assumed. This has immediate implications for understanding recent electoral surprises, platform governance decisions, and primary system design.

Our analysis also reveals the previously unrecognized importance of network centrality in determining democratic outcomes, showing that the partisan identity of highly connected actors can matter more than aggregate population characteristics or individual voter sophistication. Independent voters in central network positions function as ``epistemic circuit breakers" that prevent partisan echo chambers from becoming completely isolated from reality, while strategic partisans in those same positions can amplify bias throughout the entire system by broadcasting systematically distorted signals to large audiences across multiple network clusters. Our network-centric approach resonates with broader trends in political science that emphasize how structural interdependencies shape political outcomes \citep{ward2011network}, how social contexts condition individual political behavior \citep{sinclair2012social}, and how understanding interdependence among actors represents a fundamental challenge for social science research. Reform should focus on who controls hub positions and on how algorithms rank political voices, not only on voter training.

The practical implications for democratic reform are both urgent and actionable. Our framework suggests that platform designers should prioritize metrics beyond engagement, incorporating measures of source credibility, political independence, and accuracy when determining content visibility. Rather than treating all political speech as equivalent, platforms might systematically amplify voices that serve epistemic circuit-breaking functions while limiting the reach of communications exhibiting strategic bias patterns. Regulators need metrics that flag extreme voices and track how fast their messages spread. Timing matters. Long debate windows can let bias grow, so institutions might limit rounds or stagger information release.

Looking toward future research, our framework opens multiple avenues for theoretical development and empirical validation. Next steps include testing the model in multi-candidate races, allowing ties to form endogenously, and running lab or field experiments on platform tweaks. Future work should examine how these mechanisms operate across different democratic institutions and electoral systems, how elite responsiveness to public opinion \citep{sevenans2021public} might be shaped by the same network dynamics that affect mass political behavior, and how our theoretical framework can inform broader questions about representation and democratic accountability in networked societies. Perhaps most importantly, our approach suggests broader applications beyond electoral politics to any collective decision context where groups must choose based on uncertain information while members maintain conflicting interests—from corporate governance and expert panels to jury deliberation and international negotiations. By demonstrating how communication network architecture shapes collective intelligence, our analysis provides tools for designing institutions that harness the benefits of social learning while defending against the systematic biases that can lead democratic societies away from truth and toward collectively harmful choices.

\section{Conclusion}

Classic democratic theory sees networks as pipes of collective wisdom, letting diverse views merge into better group judgments. Our analysis fundamentally challenges this optimistic vision by revealing how the same network structures that can enhance democratic deliberation through noise filtering can simultaneously undermine it through partisan bluffing that systematically distorts collective beliefs. Networks embody a fundamental tension between information aggregation and bias amplification, resolving puzzles about when deliberation improves versus degrades group decision-making. We show that mid-level competition is most fragile: when quality gaps match partisan bias in size, networks tip toward error, inverting the standard competition–quality link.

Our results shift blame from voter ignorance to network design, showing that architecture tilts the balance between truth and bias. The discovery that independent actors in central network positions function as ``epistemic circuit breakers" suggests concrete strategies for preserving democratic competence: platforms can boost credible non-partisan voices, schedule forced cross-group exposure near elections, and track the reach of extreme rhetoric, not just its volume. As artificial intelligence systems increasingly mediate political communication, design choices made now will lock in either a corrective or corrosive information equilibrium for years. The choice between these futures depends critically on whether democratic societies can design communication technologies that harness the benefits of networked deliberation while defending against its systematic pathologies.

The broader implications extend beyond electoral politics to any collective decision context where groups choose based on uncertain information while maintaining conflicting interests. Our framework provides tools for analyzing communication dynamics in corporate governance, expert panels, jury deliberation, and international negotiations. Because bias compounds but random noise cancels, designers should keep honest hubs, foster cross-cutting ties, and pace deliberation strategically. The stakes transcend academic debates to encompass the foundational challenge of governing complex societies through collective choice. If communication networks systematically bias democratic decisions toward collectively harmful outcomes, the legitimacy of democratic governance itself comes under strain, potentially undermining social cooperation and institutional stability. Conversely, if networks can be architected to promote accurate collective judgment while preserving pluralistic competition, they offer unprecedented tools for enhancing collective intelligence. Our study offers both caution that digital links can mislead and a path for reforms that harness them for truth. The next democratic era will hinge on building systems that prize accuracy as much as engagement and keep debate both open and resilient.

\appendix
\counterwithin{figure}{section}
\counterwithin{table}{section}

\section{Appendix}

\subsection{Conditions for Unbounded Belief Divergence in Political Networks}

\begin{figure}[!ht]
   \centering
   \caption{Conditions for Unbounded Belief Divergence in Political Networks}
   \label{fig:unbounded_belief}
   
   \begin{subfigure}{0.49\textwidth}
        \includegraphics[width=\linewidth]{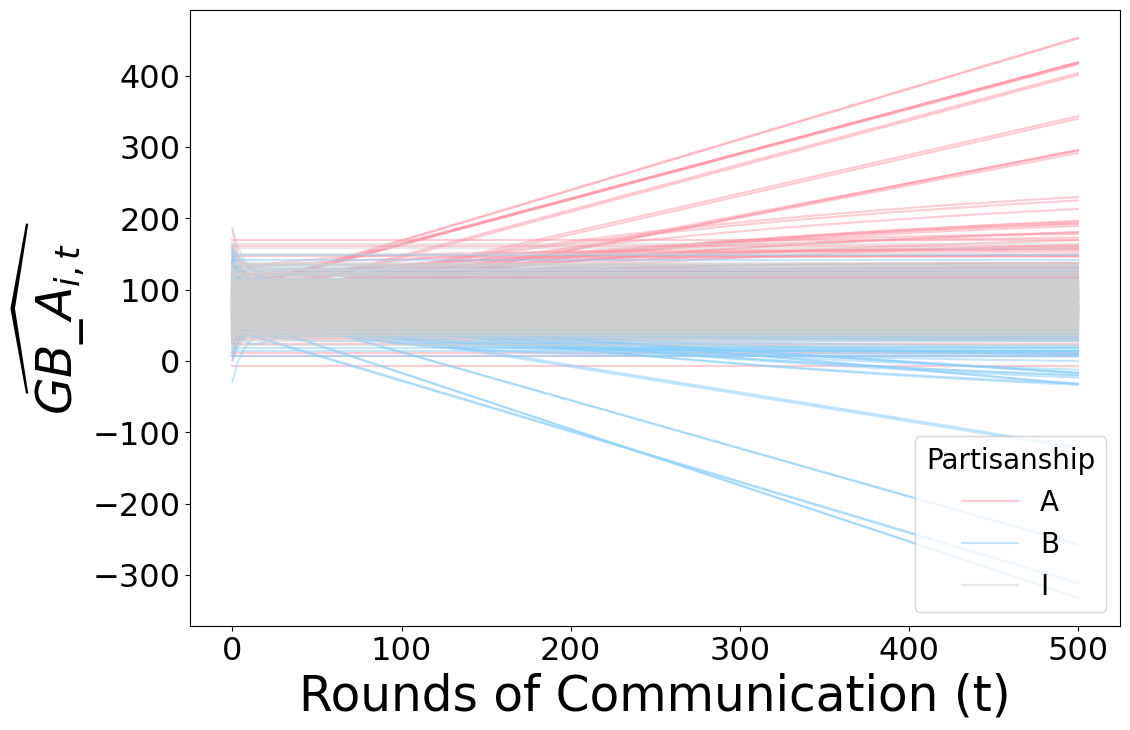}
       \caption{Low Network Density ($\langle k \rangle = 2$)}
       \label{subfig:unbounded_small_avg_degree}
   \end{subfigure}
   \hfill
   \begin{subfigure}{0.49\textwidth}
        \includegraphics[width=\linewidth]{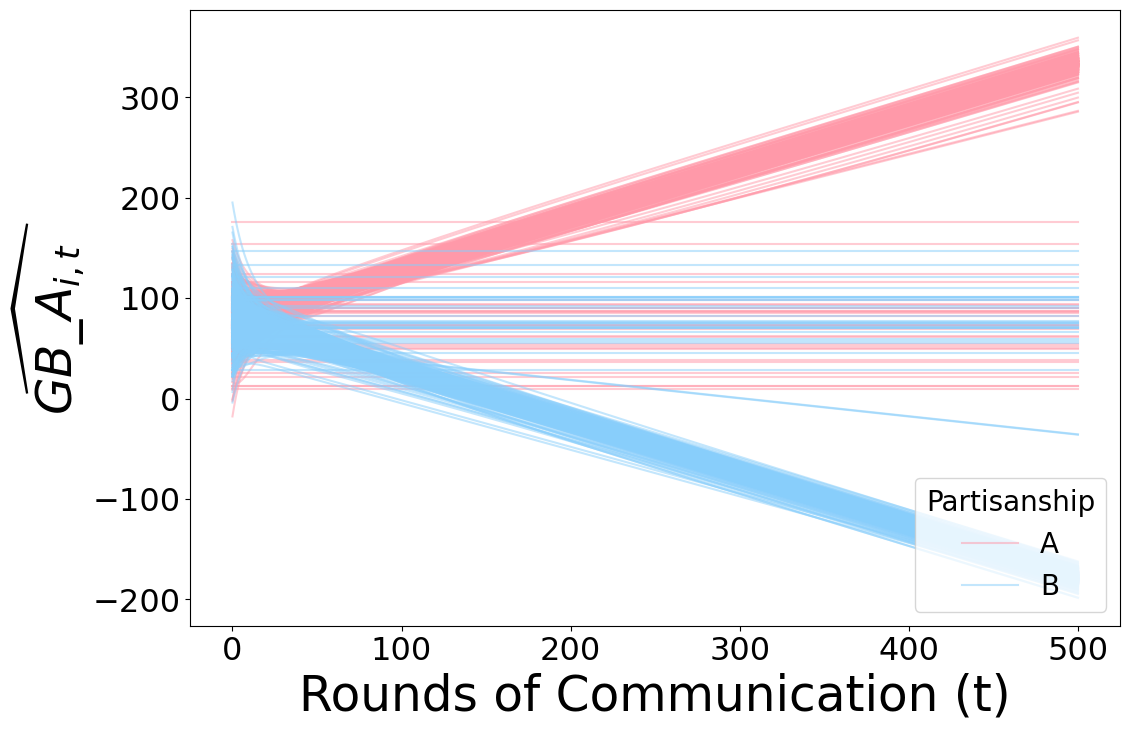}
       \caption{No Independent Voters}
       \label{subfig:unbounded_no_indep}
   \end{subfigure}
   
   \bigskip
   
   \begin{subfigure}{0.49\textwidth}
        \includegraphics[width=\linewidth]{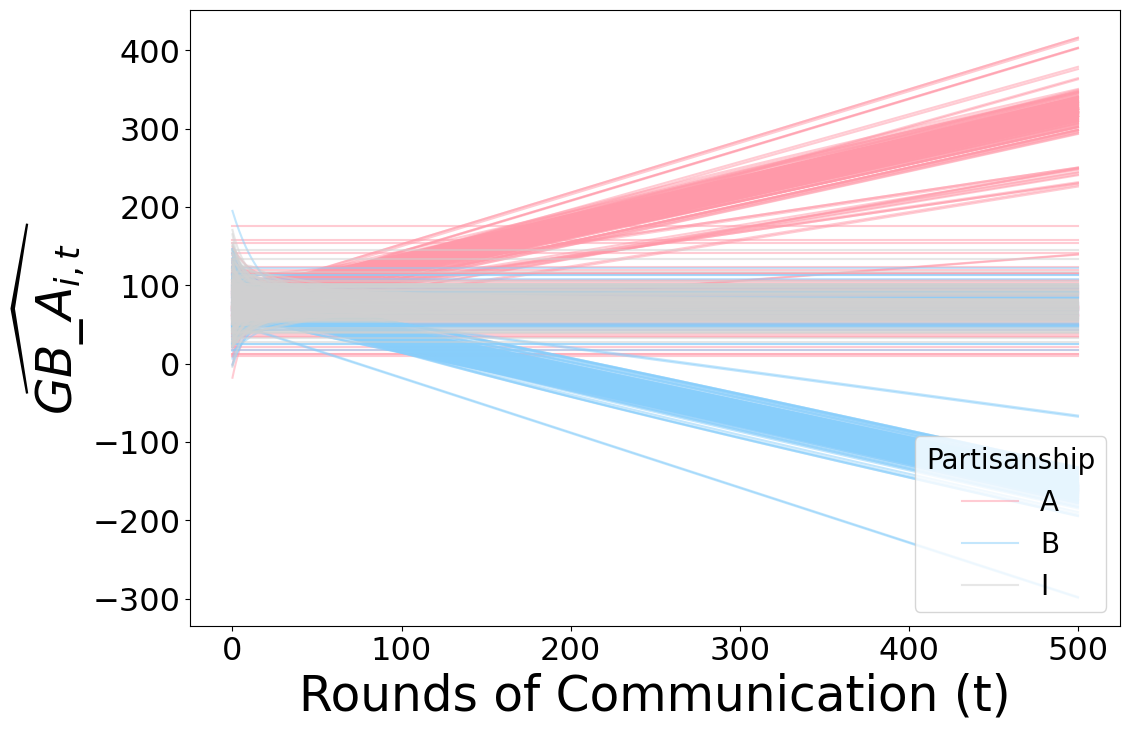}
       \caption{Partisans Ignore Independent Voices}
       \label{subfig:unbounded_partisan_no_indep}
   \end{subfigure}
   \hfill
   \begin{subfigure}{0.49\textwidth}
        \includegraphics[width=\linewidth]{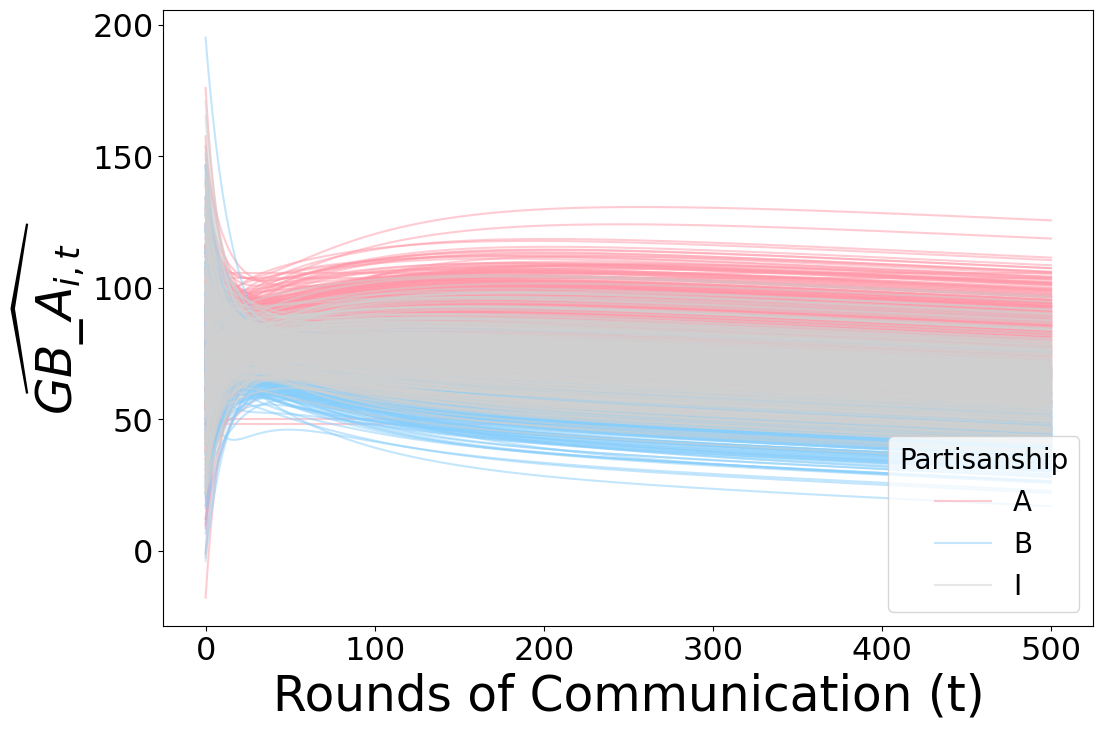}
       \caption{Independents Trust Partisan Sources}
       \label{subfig:unbounded_indep_trust_partisan}
   \end{subfigure}
\end{figure}

To further illustrate the theoretical conditions under which partisan bluffing can lead to unbounded belief divergence, we examine four alternative network configurations that systematically remove or modify the stabilizing mechanisms identified in our main analysis. These scenarios demonstrate how structural features of communication networks determine whether democratic systems converge toward bounded collective delusion or spiral toward complete detachment from reality. Figure \ref{fig:unbounded_belief} presents belief evolution patterns for candidate A quality under configurations that eliminate key circuit breakers in democratic communication.

Figure \ref{fig:unbounded_belief}(a) demonstrates how reduced network density can destabilize democratic communication by creating isolated partisan clusters with insufficient cross-cutting exposure. When average degree $\langle k \rangle$ decreases from our benchmark value of 6 to 2, the network fragments into small, homogeneous communities where partisan bluffing can operate without corrective influence from diverse information sources. The dramatic belief divergence visible in both directions—with some partisan A supporters reaching extreme positive assessments above 400 while some partisan B supporters develop extreme negative assessments below $-200$—illustrates how network fragmentation enables unbounded bias amplification within isolated echo chambers. This finding reveals that democratic systems require sufficient connectivity density to maintain epistemic coherence across different partisan communities.

The most extreme breakdown occurs in Figure \ref{fig:unbounded_belief}(b), which examines a purely partisan society with no independent voters. Under this configuration, partisan bluffing operates without any honest communication to provide corrective signals, leading to explosive belief divergence that accelerates throughout the communication process. Partisan A supporters develop increasingly inflated assessments that approach 500 by round 500, while partisan B supporters spiral toward assessments approaching $-300$, representing complete collective delusion on both sides. The symmetric but opposite trajectories demonstrate how the absence of independent ``circuit breakers" allows strategic bias to compound indefinitely through network transmission, confirming our theoretical prediction that democratic systems require some proportion of honest actors to maintain stability.

Figure \ref{fig:unbounded_belief}(c) reveals how trust structures determine democratic stability by examining scenarios where partisans completely ignore independent voices (setting $w_{p,I} = 0$ instead of our benchmark $w_{p,I} = 0.5$). Even though independents maintain accurate beliefs throughout the process (visible as stable gray lines near the true value), their inability to influence partisan belief formation eliminates their circuit-breaking function. The result resembles the purely partisan scenario, with explosive belief divergence occurring because partisans effectively isolate themselves from corrective information despite its availability in the network. This demonstrates that the mere presence of independent voters is insufficient for democratic stability—their voices must carry sufficient credibility with partisan actors to moderate extreme biases.

Finally, Figure \ref{fig:unbounded_belief}(d) examines the consequences of corrupting independent judgment by allowing independents to partially trust partisan sources (setting $w_{I,p} = 0.3$ instead of our benchmark $w_{I,p} = 0$). This configuration destroys the epistemic anchor function that independents serve in our main analysis, as their beliefs become contaminated by strategic bias transmitted through partisan networks. The result shows independents losing their stabilizing role and developing biased assessments that drift substantially from truth, while partisan groups experience amplified divergence because their traditional source of corrective information has been compromised. This scenario illustrates how democratic stability depends critically on preserving institutions and actor types that remain insulated from partisan strategic incentives.

These results provide crucial insights into the structural prerequisites for democratic stability in networked communication environments. They demonstrate that bounded collective delusion—while problematic—represents a preferable outcome to the unbounded belief divergence that occurs when key stabilizing mechanisms are removed. The findings suggest that institutional designers should focus on preserving sufficient network connectivity, maintaining populations of independent actors, ensuring that independent voices carry credibility with partisan audiences, and protecting the epistemic independence of honest communicators from strategic contamination. Without these structural safeguards, democratic communication systems can spiral toward complete detachment from reality, making accurate collective choice impossible regardless of the objective quality differences between electoral alternatives.

\subsection{Individual-Level Belief Trajectories Across Distributional Parameters}

This appendix provides detailed visualization of how individual voter beliefs evolve over time under different configurations of the partisan benefit distribution. Using our benchmark scenario where $GB_A = 77.9$ and $GB_B = 46.4$, we examine belief trajectories when partisan benefits follow a Gamma(1, $\beta$) distribution for varying values of the shape parameter $\beta$. These individual-level dynamics illuminate the precise mechanisms through which extreme partisan preferences contaminate collective belief formation and drive supporters of inferior candidates toward systematic collective delusion.

\begin{figure}[!ht]
    \centering % Center the figure in the text area

    \caption{Individual Belief Evolution Under Alternative Partisan Benefit Distributions}
    \label{fig:PB_belief}

    \captionsetup[subfigure]{justification=centering, font=bf, size=small, singlelinecheck=false}
    
    \begin{subfigure}{0.49\textwidth}
        \includegraphics[width=\linewidth]{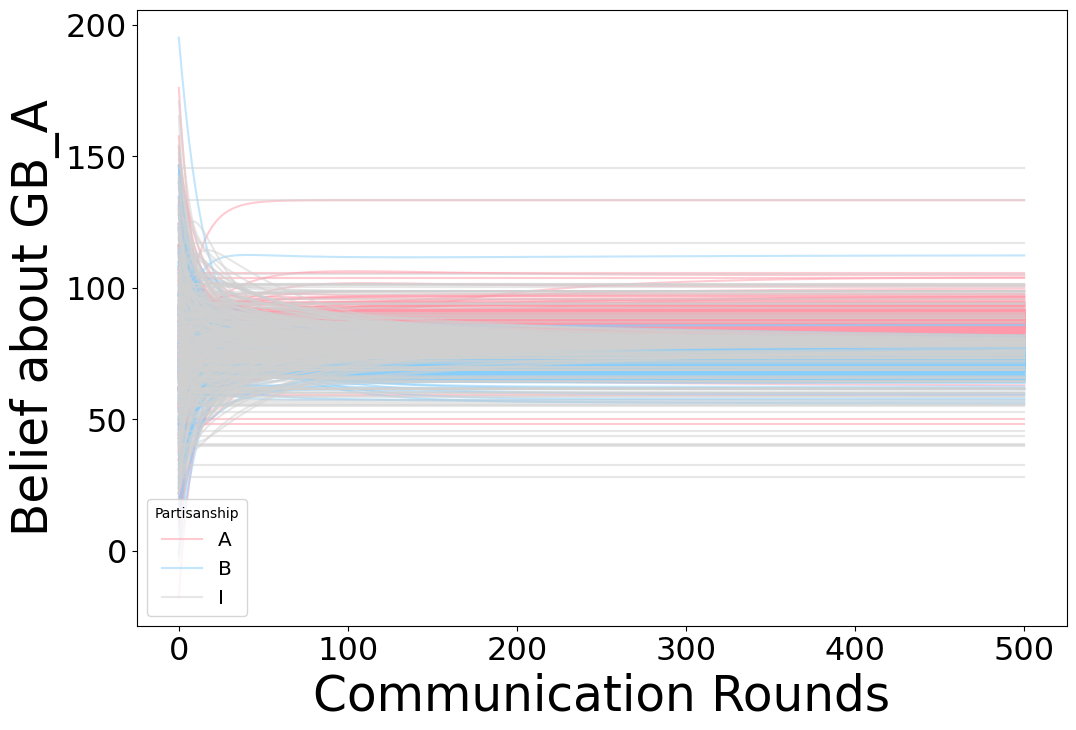}
        \caption{\( \beta \) = 0.4}
    \end{subfigure}
    \hfill 
    \begin{subfigure}{0.49\textwidth}
        \includegraphics[width=\linewidth]{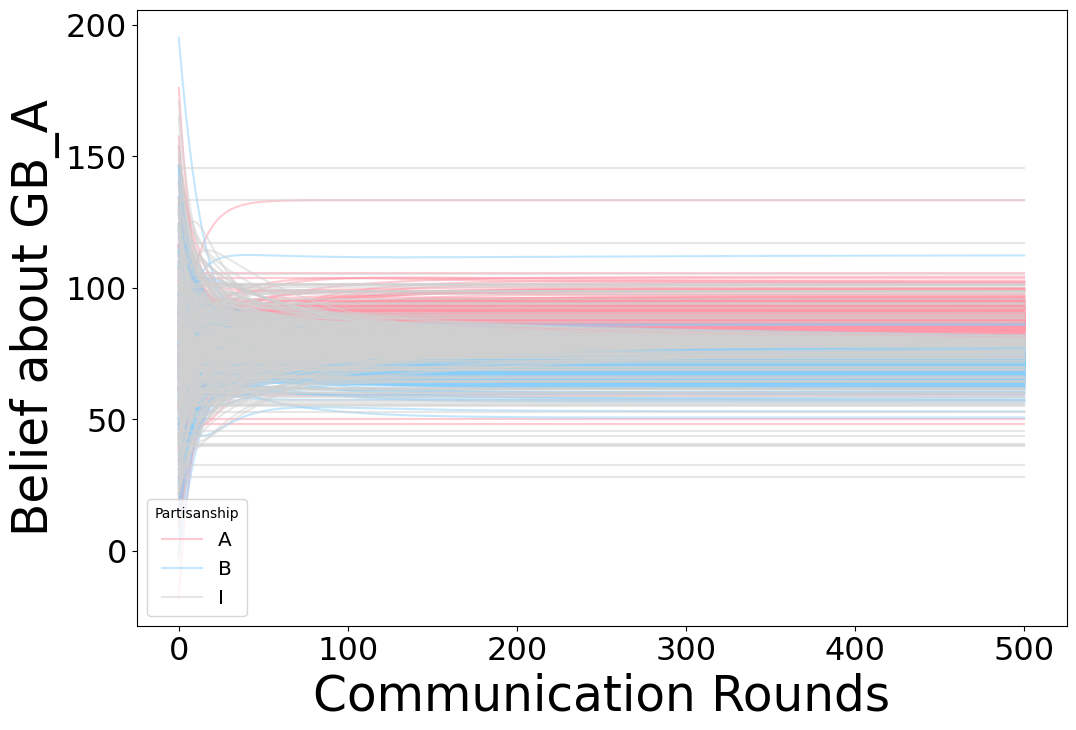}
        \caption{\( \beta\) = 0.3}
    \end{subfigure}

    \bigskip % This adds some vertical space between the rows of figures

    \begin{subfigure}{0.49\textwidth}
        \includegraphics[width=\linewidth]{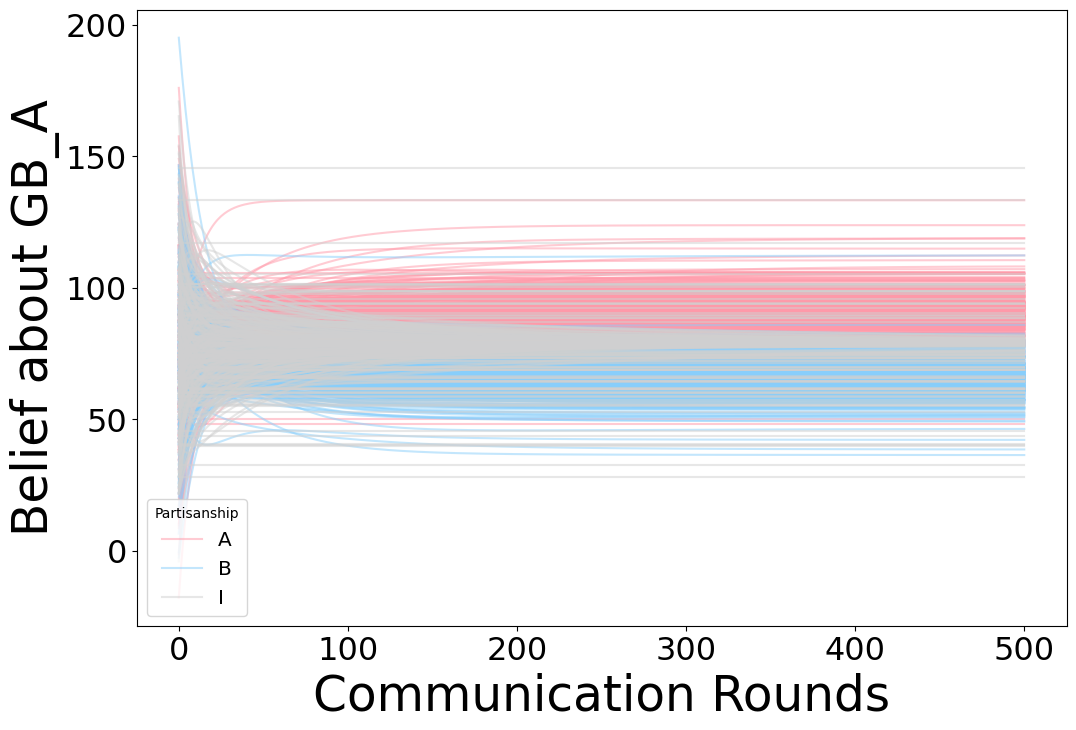}
        \caption{\( \beta\) = 0.2}
    \end{subfigure}
    \hfill % This will add space between the two figures
    \begin{subfigure}{0.49\textwidth}
        \includegraphics[width=\linewidth]{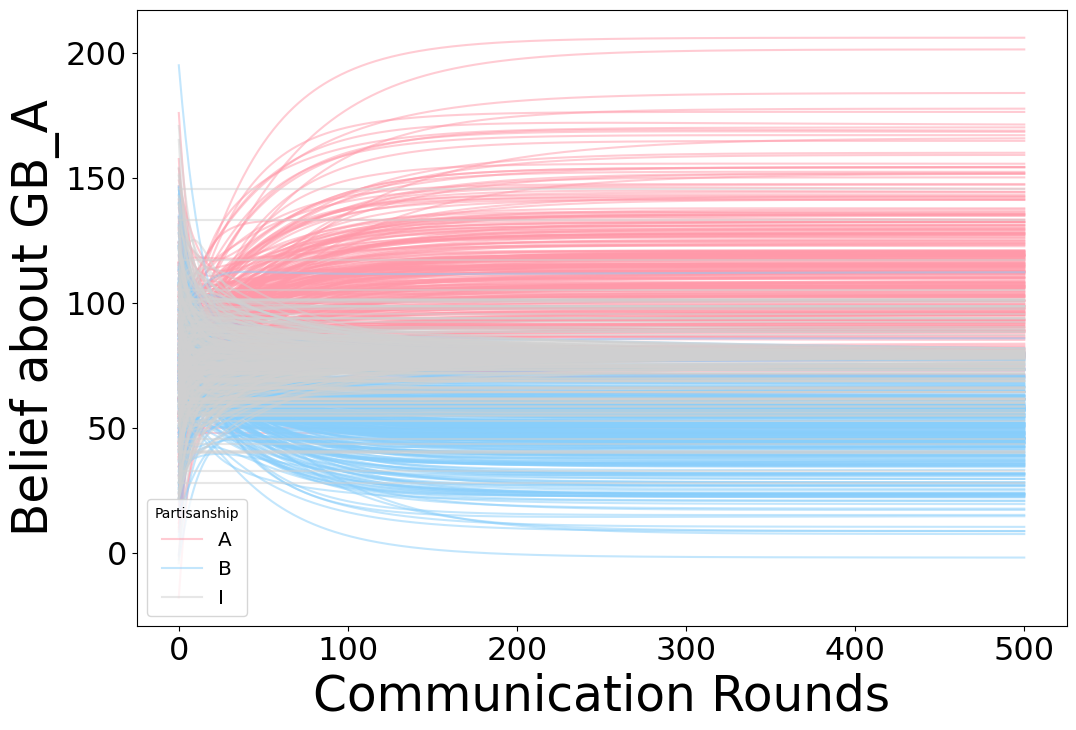}
        \caption{\( \beta\) = 0.1}
    \end{subfigure}

    % \footnotesize Notes:

\end{figure}

Figure \ref{fig:PB_belief} presents belief trajectories for all 1,000 voters across four different distributional configurations, with each line representing an individual voter's assessment of candidate A quality over 500 communication rounds. The color coding distinguishes partisan A supporters (red), partisan B supporters (blue), and independents (gray), revealing how different distributional assumptions generate dramatically different patterns of belief convergence and divergence.

The progression across panels reveals the profound influence of distributional tail behavior on democratic communication dynamics, demonstrating how increasing prevalence of extreme partisans systematically degrades collective belief formation. Figure \ref{fig:PB_belief} (a) presents the baseline case where $\beta = 0.4$ creates a highly concentrated distribution with minimal extreme preferences. Under this configuration, belief trajectories for all voter types converge toward truth with minimal systematic bias, demonstrating that democratic communication can achieve its theoretical potential when strategic incentives remain moderate. The tight clustering of individual trajectories around true values illustrates how concentrating partisan preferences enables beneficial information aggregation to dominate harmful bias amplification.

Figure \ref{fig:PB_belief} (b) reveals the initial signs of democratic degradation as $\beta = 0.3$ introduces a longer distributional tail with more extreme partisan preferences. While most belief trajectories remain relatively stable, we begin to observe systematic bias among partisan B supporters with increased variance across individual voters. The systematic bias among partisan B supporters becomes visible but remains bounded, suggesting that moderate increases in extreme preferences can limit the scope of accurate collective judgment without creating complete collective delusion. Independent voters (gray lines) maintain consistent trajectories across all distributional configurations, confirming their role as epistemic anchors that provide stable reference points regardless of partisan dynamics.

The deterioration accelerates in Figure \ref{fig:PB_belief} (c), where $\beta = 0.2$ creates a moderate tail distribution that substantially amplifies bias effects without completely destabilizing the system. Partisan B supporters exhibit systematic overestimation of candidate A quality, with the range of individual trajectories widening considerably compared to the concentrated distribution scenario, and the rate of belief divergence increasing noticeably. This intermediate case reveals the nonlinear relationship between distributional parameters and democratic performance: modest increases in the prevalence of extreme partisans can generate large deteriorations in collective accuracy through cascading effects in network communication.

Finally, Figure \ref{fig:PB_belief} (d) demonstrates the complete breakdown of democratic communication when $\beta = 0.1$ creates a heavy-tailed distribution with numerous extreme partisans. We observe dramatic belief polarization that persists throughout the entire communication process, with partisan B supporters (blue lines) exhibiting severe upward drift in their assessments of candidate A, many individual trajectories reaching assessments above 150 despite the true value of 77.9. This systematic overestimation reflects the amplified influence of extreme partisan B voters who transmit heavily biased signals that contaminate belief formation throughout their network neighborhoods. The wide dispersion of belief trajectories among partisan B supporters indicates that extreme preferences create heterogeneous bias amplification, with voters embedded in different network configurations experiencing varying degrees of collective delusion.

These individual-level dynamics provide crucial evidence for our theoretical framework by demonstrating how certain voters with extreme partisan benefits function as ``bias super-spreaders" who contaminate collective belief formation far beyond their numerical representation. The visualization reveals that extreme partisans do not simply hold biased beliefs in isolation—they actively transmit these biases through network communication, creating cascading effects that shift entire partisan communities toward collective delusion. The progressive amplification of bias effects as distributional tails expand confirms that democratic vulnerability depends critically on the prevalence of extreme preferences rather than just the mean level of partisan attachment.

Moreover, the persistence of biased trajectories across hundreds of communication rounds demonstrates that strategic communication creates stable equilibria that resist correction through continued social interaction. Unlike random errors that cancel through information aggregation, the systematic biases generated by extreme partisans compound through network transmission, leading to permanent departures from truth that persist indefinitely. This finding challenges optimistic assumptions about the self-correcting nature of democratic deliberation and highlights the importance of institutional mechanisms that can moderate extreme preferences or limit their influence on collective decision-making.

\subsection{Interactive Effects of Candidate Quality and Partisan Benefit Distributions}

\begin{figure}[ht]
  \centering
  \includegraphics[width=0.9\linewidth]{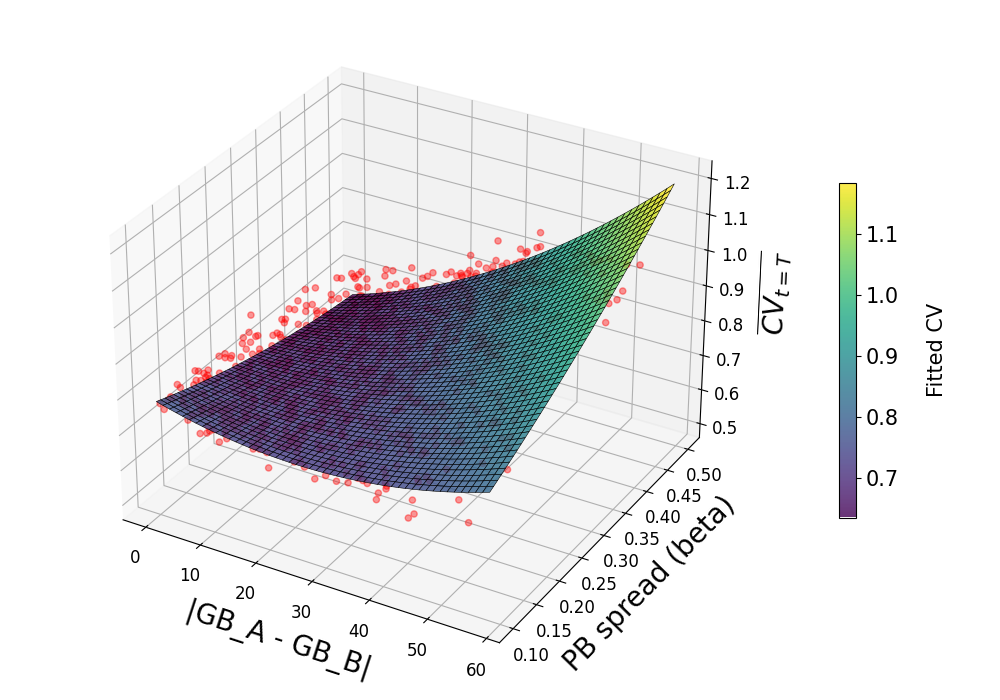}
  \caption{Interactive Effects of Candidate Quality Differences and Partisan Benefit Distributions}
  \label{fig:PB_3D}
\end{figure}

The relationship between candidate quality differences and partisan benefit distributions creates complex interaction effects that determine when democratic systems remain robust versus when they become vulnerable to systematic failure. Figure \ref{fig:PB_3D} provides a comprehensive three-dimensional analysis of these interactions based on 1,000 simulations across systematically varied combinations of global benefit differences and distributional parameters.

The surface plot reveals several critical patterns that illuminate the conditions under which democratic performance transitions between different regimes. When candidate quality differences are large (right region of the surface), democratic performance remains remarkably robust even when extreme partisans are numerous, as overwhelming objective evidence can overcome even heavily biased communication. The mathematical intuition is straightforward: when true signal strength greatly exceeds the magnitude of strategic distortion, honest information dominates partisan bluffing regardless of network composition or distributional characteristics.

However, when quality differences are moderate—precisely the range we identified as most problematic in our main analysis—the presence of extreme partisans can completely destabilize democratic decision-making, creating sharp performance cliffs where small increases in extreme preferences generate dramatic deteriorations in collective accuracy. These cliff effects occur because moderate candidate differences create optimal conditions for strategic bias to overcome truthful signals: when true quality differences become comparable to the magnitude of bias that extreme partisans introduce, network amplification can systematically distort collective beliefs away from objective reality.

The surface exhibits smooth, continuous transitions rather than abrupt discontinuities, indicating that democratic performance degrades gradually as distributional parameters shift rather than collapsing suddenly at critical thresholds. This gradual transition pattern has profound implications for understanding democratic vulnerability in contemporary societies. Rather than facing sudden systemic failures that would trigger immediate institutional responses, democratic systems may experience slowly mounting vulnerability as partisan sorting increases extreme preferences, with degradation remaining largely invisible until competitive elections reveal the extent of collective delusion among supporters of inferior candidates. This finding suggests that monitoring the distributional characteristics of partisan preferences may be as important as tracking mean levels of polarization for assessing democratic health and designing early warning systems for institutional breakdown.

\bibliographystyle{apalike}
\bibliography{ref}

\end{document}